\documentclass[12pt,draftcls,onecolumn]{IEEEtran}
\usepackage{etoolbox}
\makeatletter
\patchcmd{\@makecaption}
  {\scshape}
  {}
  {}
  {}
\makeatletter
\patchcmd{\@makecaption}
  {\\}
  {.\ }
  {}
  {}
\makeatother
\usepackage{amsmath}
 \usepackage{amsmath,amssymb}
 \usepackage{subfigure}
 \usepackage{graphicx,graphics,color,psfrag}
 \usepackage{cite,balance}
 \usepackage{cite}
 \usepackage{caption}
 \allowdisplaybreaks
 \usepackage{algorithm}
 \usepackage{accents}
 \usepackage{amsthm}
 \usepackage{bm}
 \usepackage{algorithmic}
 \usepackage[english]{babel}
 \usepackage{multirow}
 \usepackage{enumerate}
 \usepackage{cases}
 \usepackage{stfloats}
 \usepackage{dsfont}
 \usepackage{color,soul}
 \usepackage{amsfonts}
 \usepackage{cite,graphicx,amsmath,amssymb}
 \usepackage{subfigure}
 \usepackage{fancyhdr}
 \usepackage{hhline}
 \usepackage{graphicx,graphics}
 \usepackage{array,color}
 \usepackage{amsmath}
\usepackage{float}
\usepackage{amssymb}
\usepackage{amsmath}
\usepackage{amsthm}
\usepackage{amsfonts}
\usepackage{graphicx}
\usepackage{algorithm}
\usepackage{algorithmic}
\usepackage{epstopdf}
\usepackage{cite}
\usepackage{amsmath,bm}
\usepackage{subfigure}
\usepackage{graphicx}
\usepackage{color}
\usepackage{graphicx}
\usepackage{calc}
\usepackage{caption}
\usepackage{diagbox}
\makeatletter

\newcommand{\Rmnum}[1]{\expandafter\@slowromancap\romannumeral #1@}
\makeatother



\newtheorem{example}{Example}

\newtheorem{lemma}{\textbf{Lemma}}

\newtheorem{proposition}{\textbf{Proposition}}

\begin{document}
\title{Mobile Communications, Computing and Caching Resources Optimization for Coded Caching with Device Computing}
\author{Yingjiao Li, Zhiyong Chen, and Meixia Tao
\thanks{

Authors are with Shanghai Jiao Tong University, Shanghai 200240, China (e-mail: \{lyj970222, zhiyongchen, mxtao\}@sjtu.edu.cn). (\emph{Corresponding author: Zhiyong Chen}).}}
\maketitle

\begin{abstract}
Edge caching and computing have been regarded as an efficient approach to tackle the wireless spectrum crunch problem. In this paper, we design a general coded caching with device computing strategy for content computation, e.g., virtual reality (VR) rendering, to minimize the average transmission bandwidth with the caching capacity and the energy constraints of each mobile device, and the maximum tolerable delay constraint of each task. The key enabler is that because both coded data and stored data can be the data before or after computing, the proposed scheme has numerous edge computing and caching paths corresponding to different bandwidth requirement. We thus formulate a joint coded caching and computing optimization problem to decide whether the mobile devices cache the input data or the output data, which tasks to be coded cached and which tasks to compute locally. The optimization problem is shown to be 0-1 nonconvex nonsmooth programming and can be decomposed into the computation programming and the coded caching programming. We prove the convergence of the computation programming problem by utilizing the alternating direction method of multipliers (ADMM), and a stationary point can be obtained. For the coded cache programming, we design a low complexity algorithm to obtain an acceptable solution. Numerical results demonstrate that the proposed scheme provides a significant bandwidth saving by taking full advantage of the caching and computing capability of mobile devices.
\end{abstract}


\newpage
\section{INTRODUCTION}
Bandwidth saving is an eternal topic in wireless communications systems, especially in the era of shortage of wireless spectrum resource. In recent years, bandwidth requirement in the wireless network has been greatly spurred by broadband applications and services, such as the immersive panoramic virtual reality (VR) video, high definition holographic gaming, and 8K/16K ultra-high definition video \cite{update}. For example, sending full immersive VR video in 16K with H. 265 requires more than $4$ Gbit/s \cite{maga_VR}. Such modern bandwidth loads impose significant challenges to today's mobile network and has driven wireless researchers and mobile operators to find ways to keep up with people's ever-growing needs in the bandwidth for a better life.

By looking at the wireless spectrum crunch problem, we notice that differing greatly from the conventional services, e.g., phone calls and text messages, broadband applications and services highly relies on an asynchronous content reuse \cite{colors}. As such, caching popular contents at the wireless edge during off-peak time can alleviate peak-hour network congestion, i.e., saving the bandwidth\cite{caching_jsac}. In the seminal paper\cite{coded}, the coded caching scheme is proposed to relieve the traffic burden by exploiting the caching size of mobile device (MD). Meanwhile, in \cite{caching_het}, authors find that exploiting the caching resources at the relay and users can provide significant throughput gain in large-scale wireless heterogeneous networks. The above studies reveal that the \textbf{edge caching}, e.g., at a base station (BS) or the mobile devices, has been regarded as a key enabling technology in future wireless networks to tackle the wireless spectrum crunch problem.

Besides edge caching, \textbf{edge computing} can also be exploited to reduce the bandwidth requirement in the wireless network \cite{TCOM_sun, yang}. In the edge computing architecture, the network operators and the service providers can place the computing servers at the network edge, e.g., BS. Taking immersive VR application for example, with the help of, the VR video can be rendered at the edge computing server based on users field of view (FoV) so as to reduce the system latency and backhaul traffic. However, the data rate of sending immersive VR video is still very high even if we use the edge computing at BS for rendering, because the classical edge computing cannot reduce the air-interface traffic load. A better solution is to perform the on-BS rendering with the on-device rendering, proposed in our previous work \cite{TCOM_sun,TWC_sun}. In this way, by exploiting the computing capability of mobile devices, BS can offload partial data (e.g., 2D FOV) to the mobile device, and then the the mobile device performs the partial rendering, e.g., computing the 2D FOV to the 3D FOV, such that a lot of bandwidth is saved.

Inspired by this, this paper proposes a \emph{coded caching with device computing} strategy to minimize the average bandwidth consumption subject to the caching size and the energy of the MD, as well as the delay constraints. The basic idea of the propose scheme is that BS with a mobile edge computing (MEC) server can multicast the coded data of the requested tasks (e.g., FOV) to devices based on the coded cache method \cite{coded}, and the coded data can be the data before or after rendering, i.e., input or output data. What's more, the data of cached in each mobile device is either the input data or the output data. As a result, with the cooperation of the edge caching and computing at the mobile device, the coded cache with computing has more space to reduce the bandwidth consumption, which is also different from our previous work \cite{TCOM_sun,TWC_sun} and the traditional work on the coded caching \cite{coded,jinbei}.
\subsection{Related Work}

Existing works on coded caching only consider the storage of mobile devices, without considering the computing of mobile devices \cite{jinbei,decentralized,codedTIT2019}. \cite{jinbei} takes an arbitrary popularity distribution into consideration and derives a new lower bound on the transmission rate of any coded caching schemes. The authors of \cite{decentralized} propose a novel coded scheme without coordination among users after \cite{coded}. The distributed storage of multiple servers is considered with coded caching in \cite{codedTIT2019}. In addition to the coded caching, there are also several current researches for caching content at the edge of wireless networks. In \cite{caching_bs}, authors found out that when the cache size of BS is larger than a threshold, the BS density can be reduced by increasing the cache size of BS. \cite{caching_D2D} consider content caching at user devices in a wireless device-to-device (D2D) network and conclude that the cache-aided D2D networks can turn storage into bandwidth. In order to minimize the request miss ratio, a optimal bandwidth allocation scheme with the content caching placement is proposed in \cite{letter_caching_bandwidth} for a heterogeneous cellular network. Nevertheless, these articles rarely consider both the cache capability and computation capability of BS and user devices.

Generally speaking, the arrival time of the task rely on the computation scheme and hence the processing on the MEC server of the task is impacted \cite{entire}. Currently, several literary have investigated the issues like \cite{efficient, MCC}. The computational data in \cite{efficient} can be split for two calculation methods: computing locally and cloud computing. In the paper \cite{MCC}, the mobile device operates in one of the two modes when a fixed size of data need to be computed: local computing or offloading. Through those papers in the traditional MEC architecture, we find that the computation tasks are both generated at user devices. Furthermore, there are also several papers combining the traditional MEC architecture with immersive VR application. \cite{liu2018mec} optimizes the viewport rendering offloading strategy under the computation capability in the MEC server and VR device. The paper \cite{edge_vr} leverages the cache-aided and computing-aided edge server for proactive computing and caching the frames. The edge computing server is exploited in \cite{social_vr} to compute VR input data product in user device. Yet the cache capability at the user device is not considered in the above researches.

Recently, there have been some works on the caching and computing in the wireless network. In \cite{caching_mec_twc}, the optimal computing offloading and caching decisions is designed for minimizing the sum computing latency in a hybrid mobile cloud/edge computation system. However, the work of \cite{caching_mec_twc} only consider the caching and computing resource in the access point (AP).
On the other hand, our previous works \cite{TCOM_sun,TWC_sun} take full advantage of edge caching and edge computing which jointly optimizes the cache and computation policy at user devices to minimize the transmission bandwidth in the MEC system, without considering the coded caching scheme.

\subsection{Main Contributions and Paper Organization}

The main contributions of this paper are summarized as follows:
\begin{itemize}
\item \textbf{The coded caching with device computing model}: This work considers a BS with a MEC server and multiple computing-enabled and caching-aided mobile devices, thus BS has different caching and computing policies to serve the mobile devices' request. For the \emph{coded input data} transmitted by BS, the mobile device can obtain the integral data and then compute the input data locally to obtain the output data of the request tasks, i.e., the computation result. Similarly, the \emph{coded output data} is transmitted by BS and the complete output data requested is recovered without computing in the mobile device when the partial output data of tasks has been cached. In addition, BS also can select to multicast the \emph{entire input (or output) data} of one requested task to those mobile devices that the task has not been cached. Obviously, this is a more general model than that of considered in the previous works \cite{TCOM_sun,TWC_sun}.
\item \textbf{The optimal strategy for minimizing bandwidth consumption}: We further analyze that under the guarantee of quality of service (QoS), different edge caching and computing policies have different bandwidth consumption.  In this sense, this work jointly optimize the coded cache and the computation scheme to decide the MDs cache whether the input data or the output data, which tasks to be coded cached and which tasks to compute locally. We thus formulate the average bandwidth minimization problem subject to the caching size and energy constraints for the mobile device, and the latency constraint for each task.
\item \textbf{Proposed algorithm for the optimization problem}: The optimal problem is a 0-1 nonconvex nonsmooth programming problem, which is NP-hard. We thus decouple the coded cache decision and the computation decision, then reformulate the problem into a computation programming and a coded cache programming to simplify the original programming. The alternating direction method of multipliers (ADMM) algorithm is used to solve the computation programming, and we prove that the nonconvex problem can converge on monotropic program based on ADMM, therefore a stationary point of the computation programming is obtained. Finally, for the coded cache programming, the problem can be decomposed into two subproblems based on the value of the cache decision, and then an algorithm is proposed in this paper to obtain the acceptable solution.
\end{itemize}

The rest of this paper is organized as follows. Section \Rmnum{2} introduces the system model in terms of coded caching model, task request and computation model, communication model and the transmission bandwidth cost. Section \Rmnum{3} formulate the problem minimizing bandwidth, and decompose the original programming into several subproblems. Low complexity algorithms for those subproblems are proposed in Section \Rmnum{4}. Simulation results are shown in Section \Rmnum{5}. Finally the conclusion is given in Section \Rmnum{6}.

\section{SYSTEM MODEL}
\begin{figure}[t]
 \begin{center}
	\includegraphics[width=11cm,height=6cm]{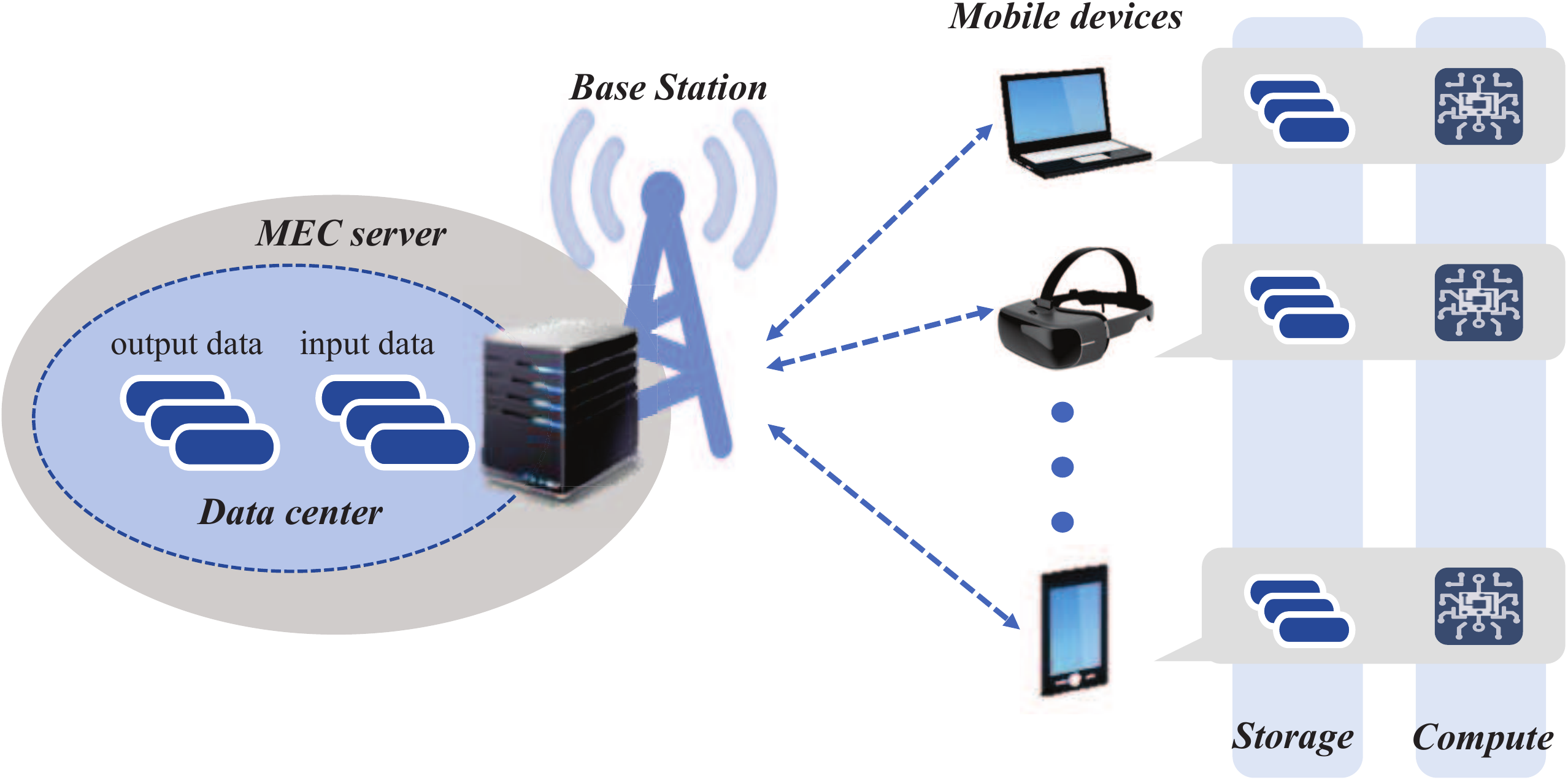}
 \end{center}
	\caption{\small{A coded caching with device computing system considered in this paper, where a BS with MEC server is connected through a wireless channel to $K$ computing-enabled and caching-aided mobile devices.} }
\label{model}
\end{figure}
Consider a general mobile downlink system with edge caching and computing, as shown in Fig.~\ref{model}, the key components include $K$ single-antenna mobile devices, denoted by a set $\mathcal{K} \triangleq \{1,2,\cdots,K\}$, and one BS with a MEC server. The MEC server is typically small-scale data center, and each mobile device can connect to the MEC via a BS. We consider the database of the computation tasks library consisting of $F$ tasks denoted as $\mathcal{F} \triangleq \{1,2,\cdots,F\}$ has been cached in the data center, e.g., there have been cached the rendered frames in the MEC server when we are watching the VR movie in the VR cinema.
\subsection{Coded Cache Model}
As shown in the model, the input data and the output data of each computation task have been cached in the MEC server. Suppose that each computation task has the same size of input data and output data, denoted as $I$(\emph{in bits}) and $O$(\emph{in bits}) respectively. We consider that each mobile device has been endowed the same capability of storage which is denoted as $C$(\emph{in bits}).

In this paper, we should determine that \emph{what kind of data is coded cached}, \emph{which task is cached} in terms of the coded cache scheme. Then, let $c_{f}$ denote the cache decision for the task $f$, and
\begin{equation}\label{equ1}
c_{f}  \in \{0,1\}, \ f\in \mathcal{F}.
\end{equation}
Here, the task $f$ is coded cached if $c_{f}=1$, and $c_{f}=0$ otherwise. Denote $\mathbf{c} \triangleq (c_{f})_{f\in \mathcal{F}}$ as the coded cache decision in the system. Let $\mathcal{N} \subseteq \mathcal{F}$ denote the set of computation tasks that are decided to be coded cached in each mobile device, i.e., $c_{n}=1$, for all $n\in \mathcal{N}$. $N = \lvert \mathcal{N} \rvert$ is the number of the tasks coded cached, and $N \leqslant K$ obviously.

In the paper, denote $d$ as the type of cached data decision for all tasks which are decided to be cached, where $d = 1$ means the type of cached data is input data, and $d = 0$ means the type of cached data is output data. In other words, the \emph{input} data of the tasks $\mathcal{N}$ is decided to be coded cached at each mobile device when $d=1$. If $d=0$, the \emph{output} data of the tasks $\mathcal{N}$ is coded cached.

According to the coded caching scheme \cite{coded}, the MEC server need to fill the storage of each mobile device based on $d$ and $\mathbf{c}$ in the cache phase. If $\mathbf{c} \ne \mathbf{0}$, i.e., $N \geqslant 1$, let $t = \lfloor \frac{CK}{NI} \rfloor$ when $d = 1$. We set $\mathcal{T} \subseteq \mathcal{K}: \lvert \mathcal{T} \rvert = t$ and denote $\mathcal{U} = \bigcup \mathcal{T}$. Therefore, the input data of the computation task $n \in \mathcal{N}$ is split into subfiles $\left(W_{n,\mathcal{T}}: \mathcal{T} \in \mathcal{U} \right)$ with equal size. During the cache phase, the MEC server transmits the subfiles $\left(W_{n,\mathcal{T}}: n \in \mathcal{N}, \mathcal{T} \in \mathcal{U}, k \in \mathcal{T} \right)$ to the mobile device $k$, and the partial input data of the tasks $\mathcal{N}$ is cached for all $k \in \mathcal{K}$. Note that there is no cached data in each mobile devices when $t = 0$ and $\mathcal{T}=\varnothing$, thus the coded cache scheme $\mathbf{c}$ is resettled to be $\mathbf{0}$. When $d = 0$, $t = \lfloor \frac{CK}{NO} \rfloor$ and the placement policy is consistent with the above.

\begin{figure}[t]
 \begin{center}
	\includegraphics[width=11cm,height=8cm]{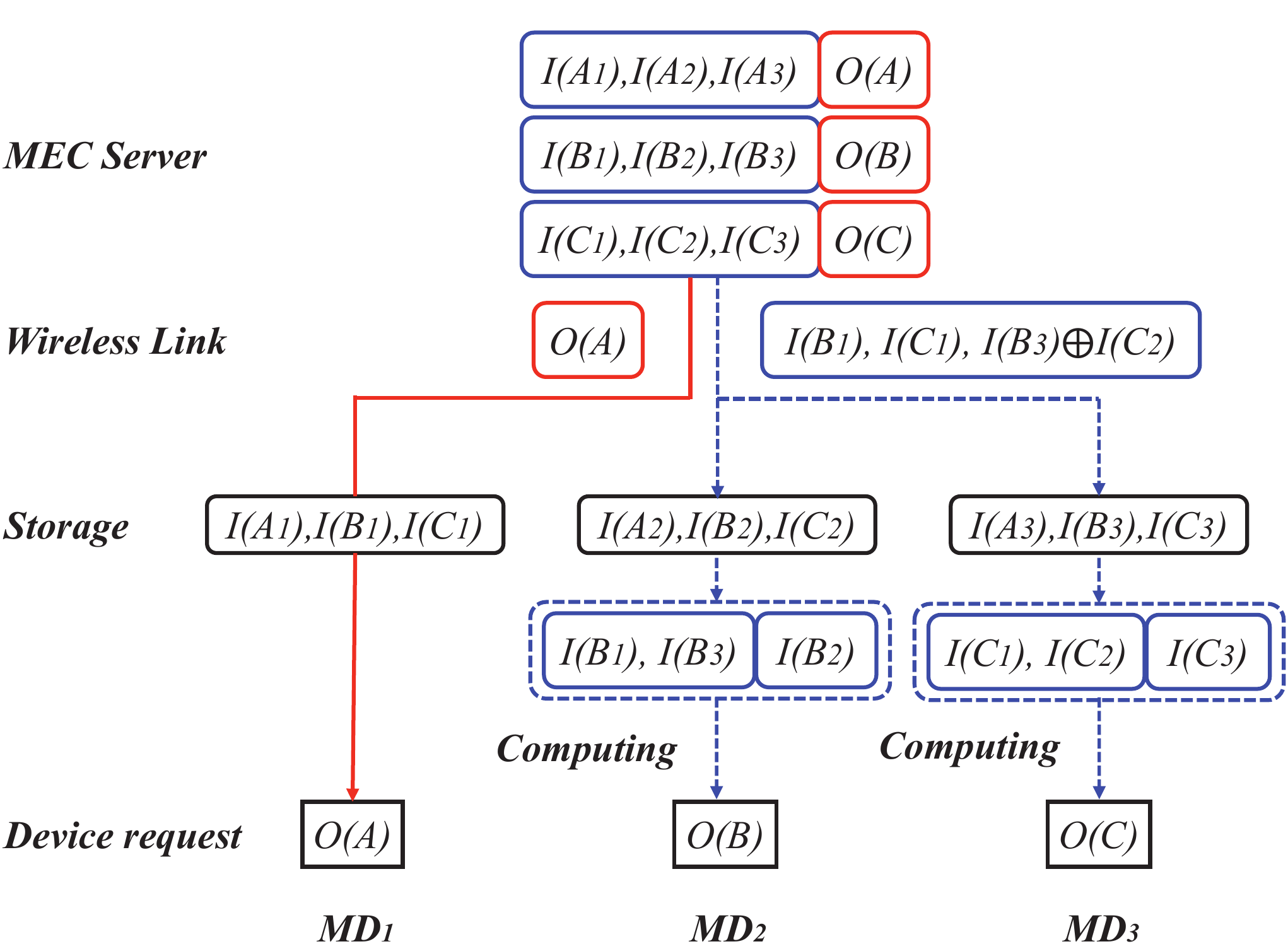}
 \end{center}
	\caption{\small{A example for $F=3$ computation tasks and $K=3$ mobile devices with the storage size $C=I$. Assume that the type of cached data is input data, i.e., $d=1$.} }
\label{example}
\end{figure}

Therefore, the computation task $n \in \mathcal{N}$ that decided to be coded cached is divided into $\binom{K}{t}$ subfiles that have no overlap with each other, then the corresponding $\binom{K-1}{t-1}$ subfiles are cached in each mobile device when $t \geqslant 1$. The constrain of the cache size constrain is satisfied as following,
\begin{equation}\label{equ2}
N\binom{K-1}{t-1}I/{\binom{K}{t}}=NIt/K \leqslant NI\frac{CK}{NI}/{K}=C,
\end{equation}
when $d = 1$, and when $d = 0$,
\begin{equation}\label{equ3}
N\binom{K-1}{t-1}O/{\binom{K}{t}}=NOt/K \leqslant NO\frac{CK}{NO}/{K}=C.
\end{equation}
Note that $t = K$ while $t > K$ since the data of each computation task $n$ always is cached as a whole if $t > K$ which is the same as the situation $t = K$, for all $n\in \mathcal{N}$.

In other to get a better understanding, we give an example in Example~\ref{exam00}.

\begin{example}\label{exam00}
As shown in Fig.~\ref{example}, consider an example $F=K=3$ where $\mathcal{K} \triangleq \{1,2,3\}$ and $C=I$. The input data of the computation tasks $A$, $B$, $C$ is coded cached at each mobile device and $\mathcal{N}=\mathcal{F}\triangleq \{A,B,C\}$ where $N=\lvert \mathcal{F} \rvert=3$, that is $\mathbf{c}=\{1,1,1\}$ and $d=1$. We split $A$, $B$ and $C$ into three subfiles of equal size since $t = 1$, and $\mathcal{U} \triangleq \{\{1\}, \{2\},\{3\}\}$ as follows
\begin{align*}
I(A) : \{I(A_1), I(A_2), I(A_3)\},~I(B) : \{I(B_1), I(B_2), I(B_3)\},~I(C) : \{I(C_1), I(C_2), I(C_3)\}.
\end{align*}
During the cache phase, the cached subfiles in each mobile device are
\begin{align*}
MD_1 : \{I(A_1), I(B_1), I(C_1)\},~MD_2 : \{I(A_2), I(B_2), I(C_2)\},~MD_3 : \{I(A_3), I(B_3), I(C_3)\}.
\end{align*}
In the transmission phase, we consider that the device $MD_1$, $MD_2$ and $MD_3$ request the computation task $A$, $B$, and $C$ respectively in a task request state. When the computation scheme is set to be $x_{1,A}=0$, $x_{2,B}=1$ and $x_{3,C}=1$, the MEC server download the output data $O(A)$ directly to the device $MD_1$ since there is no output data of the task $A$ cached. The tasks $B$, $C$ coded cached are requested and decided to be computed locally. Therefore, the MEC server will send the coded multicast transmission to the device $MD_2$ and $MD_3$, i.e., $\mathcal{M}^q=\{2,3\}$, $d_2=B$, $d_3=C$ and $\Theta=\{\{1,2\},\{1,3\},\{2,3\}\}$. During the delivery phase, the data of the coded multicast transmission based on the coded cached scheme in the paper is
\begin{align*}
 X{\left(B,C\right)}=\{I\left(B_1\right)\}, \{I\left(C_1\right)\}, \{I\left(B_3\right) \oplus I\left(C_2\right)\}.
\end{align*}
The data is multicast to the device $MD_2$ and $MD_3$ simultaneously. Then $MD_2$ and $MD_3$ recover the input data $I(B)$ and $I(C)$ respectively and compute the input data locally to obtain the computation results $O(B)$ and $O(C)$.
\end{example}
\subsection{Task Request and Computation Model}
We consider that the system model is a time-slotted system where time is divided into slots each with a duration of $\tau$ seconds, which meet latency constraint for QoE. At the beginning of each time slot, each mobile device requests the computation result of a task simultaneously. Assume that the request latency is negligible.

Denote $p_{k,f}$ as the possibility of the computation task $f$ requested by the mobile device $k$, where $\sum_{f=1}^F p_{k,f}=1$. For all $k \in \mathcal{K}$, the probability of request is mutually independent. Denote $\mathcal{Q}\triangleq \{1,2,\cdots,F^K\}$ as the set of all possible random task requests in a time slot. Let $\mathcal{S}\triangleq (\mathbf{S}^q)_{q \in \mathcal{Q}}$ denote the random task request space, and note that $\mathbf{S}^q$ is denoted as a possible random task request state in a time slot. In addition, let $s_{k,f}^q$ denote the request action in the random task request state $\mathbf{S}^q$, where $s_{k,f}^q=1$ means that the mobile device $k$ requests the computation task $f$ in the request state $\mathbf{S}^q$, and $s_{k,f}^q=0$ otherwise. Thus, $\sum_{f=1}^F s_{k,f}^q =1$ for all $k\in \mathcal{K}$. The probability $\mathbf{P}\left( \mathbf{S}^q \right)$ of a random task request state in the time slot $\mathbf{S}^q$ can be formulated as follows,
\begin{equation}\label{equ6}
\mathbf{P}\left( \mathbf{S}^q \right)=\prod_{k\in \mathcal{K}}\left(\sum_{f\in \mathcal{F}} \left(p_{k,f}s_{k,f}^q\right)\right).
\end{equation}

Assume that the CPU frequency is fixed at each mobile device and may vary over mobile devices, which is denoted as $g_{k}$ (\emph{in cycles/s}). The workload is determined by the nature of the task itself and can be obtained through off-line measurement \cite{workload}. Without loss of the generality, the workload measured by the number of CPU cycles for processing one bit of the input data is denoted as $w_{f}$ (\emph{in cycles/bit}). Then the product $Iw_{f}$ gives CPU cycles needed to successfully execute the task $f$. Let $x_{k,f}$ denote the computational selection for the task $f$ at the mobile device $k$, where
\begin{equation}\label{equ4}
x_{k,f}=
\begin{cases}
1,& \text{the computation task $f$ is computed at mobile device $k$}, \\
0,& \text{the computation task $f$ is computed at the MEC server}.
\end{cases}
\end{equation}

Note that the mobile device $k$ needs to obtain the input data via the MEC server and computes the computation task $f$ when $x_{k,f}=1$, and when $x_{k,f}=0$, the task $f$ has been computed in the MEC server, and there is no computation consumption in the mobile device. Denote $\mathbf{x} \triangleq (x_{k,f})_{k\in \mathcal{K},f\in \mathcal{F}}$ as the system computation decision.

The dynamic power is the only power considered for the mobile execution in the paper since the dynamic power dominates. The energy per cycle is proportional to the supply voltage to the CPU \cite{energy,power}. Furthermore, the clock frequency of the CPU is approximately linear proportional to the voltage supply. Therefore, the energy per cycle can be formulated as $\alpha g_{k}^2$ in the mobile device $k$, where $\alpha$ is a constant corresponding to the hardware architecture. Therefore, there is energy consumption for computing locally. Before a time slot, the task request state is unknown. Consequently, we can only restrict the possible energy consumption in the mobile device $k$ under the computation scheme and the request probability known in advance, and the constraint can be expressed as
\begin{equation}\label{equ5}
\sum_{f=1}^F p_{k,f}\alpha g_k^2 Iw_f x_{k,f} \leq E_k,
\end{equation}
and here $E_k$ is the energy limits in the mobile device $k$.
\subsection{Communication Model}
When the mobile devices request the computation results of corresponding tasks, BS transmits the required data to the mobile devices according to the decision $d$, $\mathbf{c}$ and $\mathbf{x}$ in a request state $\mathbf{S}^q$. First, we consider $d=1$, i.e., the input data is stored in each mobile device with the coded state, so that there are three transmission cases.

1) \textbf{Case 1: \{$d=1$, $x_{k,f}=1$, $c_{f}=1$\}}.
For this case, the mobile device $k$ requests a computation task whose input data has been coded cached and decides to compute locally. Denote $\mathcal{M}^q \triangleq \{m_1^q,m_2^q,\cdots, m_{M^q}^q\}$ as the set of MDs whose request are satisfied with the case in the task request state $\mathbf{S}^q$ where $\mathcal{M}^q \subseteq \mathcal{K}$ and $M^q = \lvert \mathcal{M}^q \rvert$. The set of tasks requested by $\mathcal{M}^q$ is denoted as $\mathcal{D} \triangleq \{d_{m_1^q},d_{m_2^q},\cdots,d_{m_{M^q}^q}\}$, that is when $s_{m_{i}^q,d_{m_{i}^q}}^q=1$ for the task $d_{m_{i}^q}$ at MD $m_{i}^q$, there are $x_{m_{i}^q,d_{m_{i}^q}}=1$, $c_{d_{m_{i}^q}}=1$, and $i \in \{1,2,\cdots,M^q\}$. Due to the computing at the mobile device, the transmission scheme is different from \cite{coded}. Under the conditions that we set  $\mathcal{V}^q :(\mathcal{V}^q\subseteq \mathcal{K}, \lvert \mathcal{V}^q \rvert = t+1, \exists m^q \in \mathcal{V}^q)$ and $\Theta^q = \bigcup \mathcal{V}^q$, where $m^q \in \mathcal{M}^q$. The set of subfiles that are coded is denoted as $X{\left(d_{m_1^q},d_{m_2^q},\cdots,d_{m_{M^q}^q}\right)}: \left(\oplus_{m^q \in \mathcal{V}^q}W_{d_{m^q},\mathcal{V}^q\backslash \{m^q\}}:\mathcal{V}^q \in \Theta^q \right)$ as shown in Example~\ref{exam00} where $\oplus$ denotes bitwise XOR. Then, the coded subfiles $X{\left(d_{m_1^q},d_{m_2^q},\cdots,d_{m_{M^q}^q}\right)}$ is multicast to the mobile devices $\mathcal{M}^q$ by the MEC server for satisfying the requests simultaneously. Therefore, we have the following proposition.
\begin{proposition}\label{lem01}
Denote $b^q\left(\textbf{c},\textbf{x}\right)$ as the rate of the coded multicast transmission in the task request state $\textbf{S}^q$. $b^q\left(\textbf{c},\textbf{x}\right)=0$ when $t=K$ or $M^q = 0$, and if $t<K$,  $b^q\left(\textbf{c},\textbf{x}\right)$ can be expressed as
\begin{equation}\label{equ7}
b^q\left(\textbf{c},\textbf{x}\right) =
\begin{cases}
\left(K-t\right)/\left(1+t\right),& \text{$M^q \geqslant K-t$ }, \\
\sum_{i=1}^{M^q} \binom{K-i}{t}/\binom{K}{t},& \text{$1 \leqslant M^q < K-t$}.
\end{cases}
\end{equation}
\end{proposition}
\emph{Proof}. Proof can be seen in Appendix A.

Here, $M^q$ can be expressed as $\sum_{k=1}^K \sum_{f=1}^F c_{f}x_{k,f}s_{k,f}^q$. Therefore, the data size of the coded multicast transmission is $Ib^q\left(\textbf{c},\textbf{x}\right)$. For the mobile device $k$ with $k \in \mathcal{M}^q$, the time spent on computing locally is $\frac{Iw_{f}}{g_{k}}$. To satisfy the latency deadline $\tau$, denote $r_{k,f,q}^{1}$ as the minimum transmission rate for this case in the mobile device $k$, and we can obtain that $\frac{Ib^q\left(\mathbf{c},\mathbf{x}\right)}{r_{k,f,q}^{1}}+\frac{Iw_{f}}{g_{k}} = \tau$. Hence, $r_{k,f,q}^{1}$ can be formulated as
\begin{equation}\label{equ8}
r_{k,f,q}^{1} = Ib^q\left(\mathbf{c},\mathbf{x}\right)/\left(\tau-\frac{Iw_{f}}{g_{k}}\right)\textbf{1}(c_{f}x_{k,f}s_{k,f}^q=1),
\end{equation}
where $\textbf{1}(\cdot)$ denotes the indicator function. Suppose that the computation duration locally at each mobile device $k \in \mathcal{K}$ for the task $f$ always meets the latency constraint, i.e., $\tau \geqslant \frac{Iw_{f}}{g_{k}}$. Since the centralized content multicast is employed, we define $R_{q}^1$ as the coded multicast transmission rate, and then have
\begin{equation}\label{equ9}
R_{q}^1 \geqslant r_{k,f,q}^{1}, \forall k \in \mathcal{K}, \forall f \in \mathcal{F}.
\end{equation}

2) \textbf{Case 2: \{$d=1$, $x_{k,f}=1$, $c_{f}=0$\}}. The mobile device $k$ requests a computation task $f$ decided to computed locally. However, the computation task $f$ has not been coded cached, which is different from the previous case. BS can multicast the entire input data to the mobile devices that request the task $f$. Similarly, the minimum transmission rate $r_{k,f,q}^{2}$ for this case can be expressed as
\begin{equation}\label{equ10}
r_{k,f,q}^{2} = \frac{I}{\tau-\frac{Iw_{f}}{g_{k}}}\textbf{1}((1-c_{f})x_{k,f}s_{k,f}^q=1).
\end{equation}
Therefore, the multicast transmission rate $R_{f,q}^2$ is
\begin{equation}\label{equ11}
R_{f,q}^2 \geqslant r_{k,f,q}^{2}, \forall k \in \mathcal{K}.
\end{equation}

3) \textbf{Case 3: \{$d=1$, $x_{k,f}=0$\}}. In this case, the output data of the task $f$ is decided to be obtained by the mobile device $k$ without computing locally, which means BS transmits the output data to the mobile device $k$. Under the latency constraint, we can get
\begin{equation}\label{equ12}
r_{k,f,q}^{3} = \frac{O}{\tau}\textbf{1}((1-x_{k,f})s_{k,f}^q=1).
\end{equation}
Similarly, the multicast transmission rate $R_{f,q}^3$ for this case is
\begin{equation}\label{equ13}
R_{f,q}^3 \geqslant r_{k,f,q}^{3}, \forall k \in \mathcal{K}.
\end{equation}


For $d=0$, the output data is decided to be coded cached in the mobile devices, and three cases are considered in the paper.

4) \textbf{Case 4: \{$d=0$, $x_{k,f}=0$, $c_{f}=1$\}}. The partial output data of the computation task $f$ has been coded cached and is decided be computed at the MEC server. The mobile devices request the tasks in the case in the request state $\mathbf{S}^q$, then the coded output data is multicast to the devices. According to the transmission phase in the paper, the data size of the coded output data is $Ob^q\left(\textbf{c},\textbf{x}\right)$. We thus have
\begin{equation}\label{equ14}
r_{k,f,q}^{4} = \frac{Ob^q\left(\textbf{c},\textbf{x}\right)}{\tau}\textbf{1}(c_{f}(1-x_{k,f})s_{k,f}^q=1).
\end{equation}
Here, $M^q$ is reformulated as $\sum_{k=1}^K \sum_{f=1}^F (1-x_{k,f})c_{f}s_{k,f}^q$. We obtain the coded multicast transmission rate $R_{q}^4$ meet those requests as follows,
\begin{equation}\label{equ15}
R_{q}^4 \geqslant r_{k,f,q}^{4}, \forall k \in \mathcal{K}, \forall f \in \mathcal{F}.
\end{equation}

5) \textbf{Case 5: \{$d=0$, $x_{k,f}=0$, $c_{f}=0$\}}. Similar to \textbf{Case 3},the task $f$ in the mobile device $k$ is computed by the MEC server, and has not been coded cached. Hence, BS transmits the entire output data to the mobile device to meet the request. The minimum transmission rate is
\begin{equation}\label{equ16}
r_{k,f,q}^{5} = \frac{O}{\tau}\textbf{1}((1-c_{f})(1-x_{k,f})s_{k,f}^q=1).
\end{equation}
The multicast transmission rate $R_{f,q}^5$ for the request of the task $f$ in this case is
\begin{equation}\label{equ17}
R_{f,q}^5 \geqslant r_{k,f,q}^{5}, \forall k \in \mathcal{K}.
\end{equation}

 6) \textbf{Case 6: \{$d=0$, $x_{k,f}=1$\}}. Similar to \textbf{Case 2}, BS transmits the entire input data to the mobile device for computing locally since there are not cached input data. And we have
\begin{equation}\label{equ18}
r_{k,f,q}^{6} = \frac{I}{\tau-\frac{Iw_{f}}{g_{k}}}\textbf{1}(x_{k,f}s_{k,f}^q=1).
\end{equation}
For reducing bandwidth, when the same task $f$ is requested by multiple devices simultaneously, we define
\begin{equation}\label{equ19}
R_{f,q}^6 \geqslant r_{k,f,q}^{6}, \forall k \in \mathcal{K}
\end{equation}
as the multicast transmission rate in the case.

\section{Problem Formulation for BANDWIDTH Minimization}
Observe that, through different exploiting edge computing and caching paths, we have different bandwidth requirements on the wireless channel. Therefore, the purpose of this paper is to minimize the average bandwidth cost by optimizing the edge computing and caching policy.
\subsection{Transmission Bandwidth}
Let $h_{k}$ denote the channel gain for the mobile device $k$ and the bandwidth is $B$ (\emph{in Hz}). For the LTE/5G NR system, the transmission power spectral density is constant across the downlink system bandwidth \cite{LTE5GNR}. Let $\varepsilon_{k}$ and $n_{0}$ are the power spectral density of the transmission power and the additive white Gaussian noise, respectively. Therefore, we have the transmission rate
\begin{equation}\label{equ20}
R =B\log_2(1+\frac{\varepsilon_{k}h_k^2}{n_{0}}).
\end{equation}

In this paper, we consider that the same data requested by the mobile devices can be grouped together and served by the multicast transmission. Considering that the multicast rate is limited by the user with the worst channel condition in one multicast group, the corresponding channel conditions in the above cases are
\begin{equation}\label{equ21}
H_{q}^1= \mathop{\max}_{k\in \mathcal{K},f\in \mathcal{F}}\frac{1}{\log_2(1+\frac{\varepsilon_{k}h_k^2}{n_{0}})}\textbf{1}(c_{f}x_{k,f}s_{k,f}^q=1),
\end{equation}
\begin{equation}\label{equ22}
H_{f,q}^2= \mathop{\max}_{k\in \mathcal{K}}\frac{1}{\log_2(1+\frac{\varepsilon_{k}h_k^2}{n_{0}})}\textbf{1}((1-c_{f})x_{k,f}s_{k,f}^q=1),
\end{equation}
\begin{equation}\label{equ23}
 H_{f,q}^3= \mathop{\max}_{k\in \mathcal{K}}\frac{1}{\log_2(1+\frac{\varepsilon_{k}h_k^2}{n_{0}})}\textbf{1}((1-x_{k,f})s_{k,f}^q=1),
\end{equation}
\begin{equation}\label{equ24}
H_{q}^4 = \mathop{\max}_{k\in \mathcal{K},f\in \mathcal{F}}\frac{1}{\log_2(1+\frac{\varepsilon_{k}h_k^2}{n_{0}})}\textbf{1}(c_{f}(1-x_{k,f})s_{k,f}^q=1),
\end{equation}
\begin{equation}\label{equ25}
H_{f,q}^5 =\mathop{\max}_{k\in \mathcal{K}}\frac{1}{\log_2(1+\frac{\varepsilon_{k}h_k^2}{n_{0}})}\textbf{1}((1-c_{f})(1-x_{k,f})s_{k,f}^q=1),
\end{equation}
\begin{equation}\label{equ26}
H_{f,q}^6= \mathop{\max}_{k\in \mathcal{K}}\frac{1}{\log_2(1+\frac{\varepsilon_{k}h_k^2}{n_{0}})}\textbf{1}(x_{k,f}s_{k,f}^q=1).
\end{equation}
%
Here, define $B_{q}^I$ as the achievable bandwidth in the task request state $\mathbf{S}^q$ when the input data is decided to be coded cached, i.e., $d=1$, and the expression is as follows
\begin{equation}\label{equ27}
B_{q}^I=R_{q}^1 \times H_{q}^1 +\sum_{f\in \mathcal{F}}R_{f,q}^2 \times H_{f,q}^2 + \sum_{f\in \mathcal{F}}R_{f,q}^3 \times H_{f,q}^3.
\end{equation}
Similar to (\ref{equ27}), the corresponding bandwidth $B_{q}^O$ if $d=0$ can be expressed as
\begin{equation}\label{equ28}
B_{q}^O = R_{q}^4 \times H_{q}^4+\sum_{f\in \mathcal{F}}R_{f,q}^5 \times H_{f,q}^5 + \sum_{f\in \mathcal{F}}R_{f,q}^6 \times H_{f,q}^6.
\end{equation}

\subsection{Problem Formulation}
 Mathematically, the optimization average bandwidth minimization problem can be formulated as follows
\begin{align*}
&\mathcal{P}1:  \ \ \ \  \mathop{\min}_{\mathbf{c},d, \mathbf{x}} \ \ \ \sum_{q \in \mathcal{Q}} \mathbf{P}\left( \mathbf{S}^q \right)\left( d \times B_{q}^I + \left(1-d\right) \times B_{q}^O\right)
\nonumber\\
&\ \ \ \ \ \ \ \ \ \ \ s.t. \ \ \ \ \ d\in \{0,1\},\nonumber\\
&\ \ \ \  \ \ \ \ \ \ \ \ \ \ \ \ \ \ \ \ \left(\ref{equ1}\right),\left(\ref{equ4}\right),\left(\ref{equ5}\right),\left(\ref{equ9}\right),\left(\ref{equ11}\right),\left(\ref{equ13}\right),\left(\ref{equ15}\right),\left(\ref{equ17}\right),\left(\ref{equ19}\right),
\end{align*}
which implements a joint design of coded cache scheme and computation scheme. The objective function is the expectation of the bandwidth in a time slot, and it is easy to observe that $\mathcal{P}1$ is a 0-1 nonlinear programming problem which is intractable to derive a close-form expression. Since there are $F^K$ task request states to be considered in the objective function that will generate huge computation, we replace the task request space with the set of samples $\mathcal{N}_{\mathcal{S}}$ as an approximation for simplify where $\mathcal{N}_{\mathcal{S}} \subset \mathcal{Q}$ and $|\mathcal{N}_{\mathcal{S}}| = N_s$ \cite{sample0}, \cite{sample1}. Note that the samples are related to the request probability. Thus, problem $\mathcal{P}1$ can be reformulated as below,
\begin{align*}
&\mathcal{P}1.1:  \ \ \ \  \mathop{\min}_{\mathbf{c},d, \mathbf{x}} \ \ \ \frac{1}{N_s}\sum_{n \in \mathcal{N}_{\mathcal{S}}} \left( d \times B_{n}^I + \left(1-d\right) \times B_{n}^O\right)
\nonumber\\
&\ \ \ \ \ \ \ \ \ \ \ s.t. \ \ \ \ \ d\in \{0,1\},\nonumber\\
&\ \ \ \  \ \ \ \ \ \ \ \ \ \ \ \ \ \ \ \ \left(\ref{equ1}\right),\left(\ref{equ4}\right),\left(\ref{equ5}\right),\left(\ref{equ9}\right),\left(\ref{equ11}\right),\left(\ref{equ13}\right),\left(\ref{equ15}\right),\left(\ref{equ17}\right),\left(\ref{equ19}\right).
\end{align*}


\subsection{Decomposition of Problem $\mathcal{P}$1.1}
It is a challenge to solve $\mathcal{P}1.1$ since the objective function is nonconvex and nonsmooth although the constraints are convex. In this subsection, the decision variables $\mathbf{x}$ and $\mathbf{c}$ are decoupled and we decompose $\mathcal{P}1.1$ into several subproblems to obtain the minimal average bandwidth. At the beginning, the problem on the top level is obtaining the computation strategy $\mathbf{x}$. Suppose that there are no coded cache design, i.e., $\mathbf{c}=\textbf{0}$. Therefore, the subproblem has no business to the variable $d$ and BS offloads the entire input data or the entire output data to the mobile devices.

When $x_{k,f}=1$, BS multicasts the entire input data of the task $f$ to the mobile device $k$, then the transmission rate at mobile device $k$ is the same as $r_{k,f,q}^6$ in a task request state $\mathbf{S}^q$. Similarly, the transmission rate in the mobile device $k$ for the requested task $f$ is $r_{k,f,q}^3$ when $x_{k,f}=0$. We optimal the computation variable $\mathbf{x}$ under the following optimization
\begin{align*}
\mathcal{P}2: \ \ \ \  \mathop{\min}_{\textbf{x}} \ \ \ &\frac{1}{N_s}\sum_{n \in \mathcal{N}_{\mathcal{S}}} \left(\sum_{f \in \mathcal{F}}\left(R_{f,n}^3 \times H_{f,n}^3 +R_{f,n}^6 \times H_{f,n}^6 \right)\right)
\\
s.t. \ \ \  &\left(\ref{equ4}\right),\left(\ref{equ5}\right),\left(\ref{equ13}\right),\left(\ref{equ19}\right).
\end{align*}

Followed by obtaining a given computation design $\mathbf{x}^{\ast}$ which is the solution of $\mathcal{P}2$ , the subproblem of optimizing variable $\mathbf{c}$ and $d$ can be formulated as
\begin{align*}
\mathcal{P}3:  \ \ \ \  &\mathop{\min}_{\mathbf{c},d} \ \ \frac{1}{N_s}\sum_{n \in \mathcal{N}_{\mathcal{S}}} \left( d \times B_{n}^I + \left(1-d\right) \times B_{n}^O\right)\\
& s.t. \ \ \ \mathbf{x}=\mathbf{x}^{\ast},\\
&\ \ \ \ \ \ \left(\ref{equ1}\right),\left(\ref{equ9}\right),\left(\ref{equ11}\right),\left(\ref{equ15}\right),\left(\ref{equ17}\right).
\end{align*}

The variables $\mathbf{c}$ and $d$ are uncoupled, and we can decomposed the above problem $\mathcal{P}3$ into two subproblems based on the type of the cache decision $d$. When $d = 1$, the one of the subproblems is expressed as follows,
\begin{align*}
\mathcal{P}3.1:  \ \ \mathop{\min}_{\mathbf{c}} \ \ \ &\frac{1}{N_s}\sum_{n \in \mathcal{N}_{\mathcal{S}}} (R_{n}^1 \times H_{n}^1 +\sum_{f\in \mathcal{F}}R_{f,n}^2 \times H_{f,n}^2)\\
s.t. \ \ \ &\mathbf{x}=\mathbf{x}^{\ast},\\
&\left(\ref{equ1}\right),\left(\ref{equ9}\right),\left(\ref{equ11}\right).
\end{align*}

If we assume that the data type of coded cached in MDs is the output data, i.e., $d = 0$, the subprobelm is
\begin{align*}
\mathcal{P}3.2:  \ \ \mathop{\min}_{\mathbf{c}} \ \ \ &\frac{1}{N_s}\sum_{n \in \mathcal{N}_{\mathcal{S}}}(R_{n}^4 \times H_{n}^4 + \sum_{f\in \mathcal{F}}R_{f,n}^5 \times H_{f,n}^5)\\
s.t. \ \ \ &\mathbf{x}=\mathbf{x}^{\ast},\\
&\left(\ref{equ1}\right),\left(\ref{equ15}\right),\left(\ref{equ17}\right).
\end{align*}

In a word, the problem $\mathcal{P}1.1$ is decomposed into problem $\mathcal{P}2$, $\mathcal{P}3.1$ and $\mathcal{P}3.2$. Next, these subproblems will be solved separately and the efficient coded caching with device computing strategy is derived.
\section{EFFICIENT CODED CACHING WITH DEVICE COMPUTING STRATEGY}
\subsection{Algorithm to Solve Problem $\mathcal{P}2$}
Obviously, the constraints $\left(\ref{equ4}\right)$, $\left(\ref{equ5}\right)$, $\left(\ref{equ13}\right)$ and $\left(\ref{equ19}\right)$ are convex, but the objective function in $\mathcal{P}2$ is nonconvex and nonsmooth due to the maximum terms. Therefore, we firstly reformulate the channel condition variables $H_{f,n}^3$ and $H_{f,n}^6$ to replace these maximum terms in the objective function, and the problem reformulated can be expressed as below
\begin{align*}
\mathcal{P}2.1: ~~~~ \mathop{\min}_{\mathbf{x}} \ \ \ &\frac{1}{N_s}\sum_{n \in \mathcal{N}_{\mathcal{S}}}\sum_{f \in \mathcal{F}}\left(R_{f,n}^3\times H_{f,n}^3+R_{f,n}^6\times H_{f,n}^6\right)
\nonumber\\
 s.t. \ \ &\left(\ref{equ4}\right),\left(\ref{equ5}\right),\left(\ref{equ13}\right),\left(\ref{equ19}\right),
\end{align*}
\begin{subequations}
\begin{flalign}\label{equ32}
&H_{f,n}^3  \geqslant \frac{1}{\log_2(1+\frac{\varepsilon_{k}h_k^2}{n_{0}})}\textbf{1}(x_{k,f}s_{k,f}^n=1), \forall k \in \mathcal{K},
\end{flalign}
\begin{flalign}\label{equ33}
&H_{f,n}^6 \geqslant \frac{1}{\log_2(1+\frac{\varepsilon_{k}h_k^2}{n_{0}})}\textbf{1}((1-x_{k,f})s_{k,f}^q=1), \forall k \in \mathcal{K}.
\end{flalign}
\end{subequations}
Here we denote that $\mathbf{R}^3 = \{R_{f,n}^3\}_{f \in \mathcal{F}, n \in \mathcal{N}_{\mathcal{S}}}$, $\mathbf{R}^6 = \{R_{f,n}^6\}_{f \in \mathcal{F}, n \in \mathcal{N}_{\mathcal{S}}}$, and $\mathbf{H}^3 = \{H_{f,n}^3\}_{f \in \mathcal{F}, n \in \mathcal{N}_{\mathcal{S}}}$, $\mathbf{H}^6 = \{H_{f,n}^6\}_{f \in \mathcal{F}, n \in \mathcal{N}_{\mathcal{S}}}$.

By reformulate the channel condition variables $H_{f,n}^3$ and $H_{f,n}^6$, $\mathcal{P}2.1$ becomes smooth problem which is easier to be tackled. Moreover, the objective function in $\mathcal{P}2.1$ is a difference of convex (DC) function, and it is still an integer programming (IP) problem. There is a large body of work that utilizes similar methods to find a suboptimal solution of the IP problem, such as Branch-and-Bound \cite{bnb}, cutting plane \cite{cut} which are usually plagued with high computational complexity. In this paper, we propose to replace the binary constraints $\left(\ref{equ4}\right)$ with an equivalent set of continuous constraints firstly \cite{continuous}.
\begin{lemma}\label{lem02}
The binary set $\{0,1\}^{KF}$ can be equivalently replaced by the intersection between a box $\mathcal{S}_b$ and a nonconvex constraint as follows:
\begin{equation}
\mathbf{x}\in \{0,1\}^{KF}\Leftrightarrow\mathbf{x}\in [0,1]^{KF}\cap \{ \mathbf{x}:\sum_{f \in \mathcal{F}}\sum_{k \in \mathcal{K}}x_{k,f}\left(1-x_{k,f}\right) \le 0 \},
\end{equation}
where the box is $\mathcal{S}_b=[0,1]^{KF}$.
\end{lemma}

We add the equivalent continuous nonconvex constraint into objective as penalty methods do \cite{penalty00}, \cite{penalty01}, \cite{penalty02} and the penalty parameter is defined as $\beta$. As far as $\mathcal{P}2.1$ is concerned, the numbers of the variables and the constraints reach $KF$ and $4N_sKF+KF+K$. Generally speaking, ADMM is always exploited to optimize large-scale convex programming, however, it has also been proved that the nonconvex problem can be tackled \cite{admm01}. In this paper, we employ ADMM to solve the problem by introduce a set of local copies of the variables $\hat{\mathbf{R}}^3$, $\hat{\mathbf{R}}^6$, and $\hat{\mathbf{H}}^3$, $\hat{\mathbf{H}}^6$ and $\mathbf{y}$, where are defined as $\hat{\mathbf{R}}^3=\{\hat{R}_{f,n}^3\}_{f \in \mathcal{F}, n \in \mathcal{N}_{\mathcal{S}}}$, $\hat{\mathbf{R}}^6=\{\hat{R}_{f,n}^6\}_{f \in \mathcal{F}, n \in \mathcal{N}_{\mathcal{S}}}$, $\hat{\mathbf{H}}^3=\{\hat{H}_{f,n}^3\}_{f \in \mathcal{F}, n \in \mathcal{N}_{\mathcal{S}}}$, $\hat{\mathbf{H}}^6=\{\hat{H}_{f,n}^6\}_{f \in \mathcal{F}, n \in \mathcal{N}_{\mathcal{S}}}$, and $\mathbf{y}=\{y_{k,f}^n\}_{k \in \mathcal{K}, f \in \mathcal{F}, n \in \mathcal{N}_{\mathcal{S}}}$ to product consensus constraints. Based on the consensus constraints, the coupling constraints $\left(\ref{equ13}\right)$, $\left(\ref{equ19}\right)$, $(\ref{equ32})$, $(\ref{equ33})$ can be rewritten as
\begin{subequations}
\begin{flalign}\label{equ49}
&\hat{R}_{f,n}^3 \geqslant \frac{I}{\tau-\frac{Iw_{f}}{f_{k}}}\textbf{1}\left(y_{k,f}^{n}s_{k,f}^n=1\right), \forall k \in \mathcal{K},
\end{flalign}
\begin{flalign}\label{equ50}
&\hat{R}_{f,n}^6 \geqslant \frac{O}{\tau}\textbf{1}\left((1-y_{k,f}^{n})s_{k,f}^n=1\right), \forall k \in \mathcal{K},
\end{flalign}
\begin{flalign}\label{equ51}
&\hat{H}_{f,n}^3  \geqslant \frac{1}{\log_2(1+\frac{\varepsilon_{k}h_k^2}{n_{0}})}\textbf{1}(y_{k,f}^{n}s_{k,f}^n=1), \forall k \in \mathcal{K},
\end{flalign}
\begin{flalign}\label{equ52}
&\hat{H}_{f,n}^6 \geqslant \frac{1}{\log_2(1+\frac{\varepsilon_{k}h_k^2}{n_{0}})}\textbf{1}((1-y_{k,f}^{n})s_{k,f}^q=1), \forall k \in \mathcal{K}.
\end{flalign}
\end{subequations}

As a result, we can obtain the equivalent version of $\mathcal{P}2.1$ given by
\begin{equation*}
\begin{aligned}
\mathcal{P}2.2:  \mathop{\min}_{\mathbf{x},\mathbf{y},\Pi, \hat{\Pi}} \ \ \ &\frac{1}{N_s}\sum_{n \in \mathcal{N}}\sum_{f \in \mathcal{F}}\left(R_{f,n}^3\times H_{f,n}^3+R_{f,n}^6\times H_{f,n}^6\right)+\beta\sum_{f \in \mathcal{F}}\sum_{k \in \mathcal{K}}x_{k,f}\left(1-x_{k,f}\right)\\
&\ \ \ \ \ \ \ \ \ \ \ \ \  s.t.\ \ \ \ \left(\ref{equ4}\right),\left(\ref{equ5}\right),\left(\ref{equ49}\right),\left(\ref{equ50}\right),\left(\ref{equ51}\right),\left(\ref{equ52}\right),\\
&\ \ \ \ \ \ \ \ \ \ \ \ \ \ \ \ \ \ \ \ \ \ y_{k,f}^n=x_{k,f}, \forall n \in \mathcal{N}_{\mathcal{S}},\\
&\ \ \ \ \ \ \ \ \ \ \ \ \ \ \ \ \ \ \ \ \ \hat{R}_{f,n}^3 = R_{f,n}^3, \forall f \in \mathcal{F}, \forall n \in \mathcal{N}_{\mathcal{S}},\\
&\ \ \ \ \ \ \ \ \ \ \ \ \ \ \ \ \ \ \ \ \ \hat{R}_{f,n}^6 = R_{f,n}^6, \forall f \in \mathcal{F}, \forall n \in \mathcal{N}_{\mathcal{S}},\\
&\ \ \ \ \ \ \ \ \ \ \ \ \ \ \ \ \ \ \ \ \ \hat{H}_{f,n}^3 = H_{f,n}^3, \forall f \in \mathcal{F}, \forall n \in \mathcal{N}_{\mathcal{S}},\\
&\ \ \ \ \ \ \ \ \ \ \ \ \ \ \ \ \ \ \ \ \ \hat{H}_{f,n}^6 = H_{f,n}^6, \forall f \in \mathcal{F}, \forall n \in \mathcal{N}_{\mathcal{S}},
\end{aligned}
\end{equation*}
where $\Pi = \{\mathbf{R}^3,\mathbf{R}^6,\mathbf{H}^3,\mathbf{H}^6\}$ and
$\hat{\Pi} = \{\hat{\mathbf{R}}^3,\hat{\mathbf{R}}^6,\hat{\mathbf{H}}^3,\hat{\mathbf{H}}^6\}$.  Moreover, we can get the following Lemma according to Theorem 5 and Theorem 8 in \cite{penalty}.

\begin{lemma}
$\mathcal{P}2.1$ and $\mathcal{P}2.2$ have the same optimal solution when the penalty parameter $\beta$ is large enough.
\end{lemma}

According to \cite{admm00}, the augmented Lagrangian function can be formulated as
\begin{equation}
\begin{aligned}\label{equl}
\mathcal{L}_{\rho}\left(\mathbf{x},\mathbf{y},\mathbf{\lambda}, \mathbf{z},\Pi,\hat{\Pi} \right) =& \Psi\left(\mathbf{x},\mathbf{y},\Pi,\hat{\Pi},\beta \right)+\frac{\rho_0}{2}\sum_{n \in \mathcal{N}_{\mathcal{S}}}\lVert\mathbf{x}-\mathbf{y}^n+\mathbf{\lambda}^n\rVert_2^2\\
&+\frac{\rho_1}{2}\lVert\hat{\mathbf{R}}^3-\mathbf{R}^3+\mathbf{z}^1\rVert_2^2+\frac{\rho_2}{2}\lVert\hat{\mathbf{R}}^6-\mathbf{R}^6+\mathbf{z}^2\rVert_2^2\\
&+\frac{\rho_3}{2}\lVert\hat{\mathbf{H}}^3-\mathbf{H}^3+\mathbf{z}^3\rVert_2^2+\frac{\rho_4}{2}\lVert\hat{\mathbf{H}}^6-\mathbf{H}^6+\mathbf{z}^4\rVert_2^2,
\end{aligned}
\end{equation}
where $\Psi\left(\mathbf{x},\mathbf{y},\Pi,\hat{\Pi},\beta \right)$ is the objective function in the problem $\mathcal{P}2.2$, and $\rho = \{\rho_0,\rho_1,\rho_2,\rho_3,\rho_4\}$ are positive penalty parameters. Note that $\mathbf{y}^n = \{y_{k,f}^n\}_{k \in \mathcal{K},f \in \mathcal{F}}$, $\mathbf{\lambda}^n=\{\lambda_{k,f}^n\}_{k \in \mathcal{K},f \in \mathcal{F}}$,  $\mathbf{\lambda}=\{\lambda^n\}_{n \in \mathcal{N}_{\mathcal{S}}}$ and $\mathbf{z}=\{\mathbf{z}^1,\mathbf{z}^2,\mathbf{z}^3,\mathbf{z}^4\}$ indicates dual variables, where $\mathbf{z}^1=\{z_{f,n}^1\}_{f \in \mathcal{F},n \in \mathcal{N}_{\mathcal{S}}}$, $\mathbf{z}^2=\{z_{f,n}^2\}_{f \in \mathcal{F},n \in \mathcal{N}_{\mathcal{S}}}$, $\mathbf{z}^3=\{z_{f,n}^3\}_{f \in \mathcal{F},n \in \mathcal{N}_{\mathcal{S}}}$, $\mathbf{z}^4=\{z_{f,n}^4\}_{f \in \mathcal{F},n \in \mathcal{N}_{\mathcal{S}}}$. Following the ADMM process, we update the primal variables $(\mathbf{x},\mathbf{y},\Pi,\hat{\Pi})$ by minimizing the augmented Lagrangian function and perform gradient ascent on the dual problem to update $\{\lambda, \mathbf{z}\}$ in each iteration. In the $t+1$ iteration, the update steps as follows.
\begin{itemize}
\item \emph{ Update the introduced variables $\{\mathbf{y}, \hat{\Pi}\}$}. Based on the previous iteration $t$, $\{\mathbf{x}(t), \mathbf{z}(t), \Pi(t)\}$ have been updated. Update $\{\mathbf{y}, \hat{\Pi}\}$ at the iteration $t+1$ by solving the following problem.
\begin{equation}
\begin{aligned}\label{equ40}
\{\mathbf{y}(t), \hat{\Pi}(t+1)\} \gets \mathop{\arg} \mathop{\min}_{\mathbf{y}, \hat{\Pi}}\mathcal{L}_{\rho}\left(\mathbf{x}(t),\mathbf{y},\mathbf{\lambda}(t), \mathbf{z}(t), \Pi(t),\hat{\Pi} \right)
\end{aligned}
\end{equation}
\item \emph{ Update the global variables $\{\mathbf{x},\Pi\}$}. Given $\{\mathbf{y}(t+1), \hat{\Pi}(t+1)\}$ obtained by solving the above problem, $\{\mathbf{x},\Pi\}$ is updated rely on the solution of the following problem.
\begin{equation}
\begin{aligned}\label{equ41}
\{\mathbf{x}(t+1), {\Pi}(t+1)\} \gets \mathop{\arg} \mathop{\min}_{\mathbf{x}, {\Pi}}\mathcal{L}_{\rho}\left(\mathbf{x},\mathbf{y}(t+1),\mathbf{\lambda}(t), \mathbf{z}(t), \Pi,\hat{\Pi}(t+1) \right)
\end{aligned}
\end{equation}
\item \emph{ Update the dual variables $\{\mathbf{\lambda}, \mathbf{z}\}$}. Depending on $\{\mathbf{x}(t+1), \mathbf{y}(t+1), {\Pi}(t+1), \hat{\Pi}(t+1)\}$, the dual variables $\{\mathbf{\lambda}, \mathbf{z}\}$ are updated as follow.
\begin{subequations}
\begin{flalign}\label{equ43}
&\mathbf{\lambda}^{n}(t+1) \gets \mathbf{\lambda}^{n}(t) + \big(\mathbf{x}(t+1)-\mathbf{y}^{n}(t+1)\big), n \in \mathcal{N},
\end{flalign}
\begin{flalign}\label{equ44}
&{\mathbf{z}^1}(t+1) \gets {\mathbf{z}^1}(t) + \left({\hat{\mathbf{R}}^3}(t+1)-{\mathbf{R}^3}(t+1)\right),
\end{flalign}
\begin{flalign}\label{equ45}
&{\mathbf{z}^2}(t+1) \gets {\mathbf{z}^2}(t) + \left({\hat{\mathbf{R}}^6}(t+1)-{\mathbf{R}^6}(t+1)\right),
\end{flalign}
\begin{flalign}\label{equ46}
&{\mathbf{z}^3}(t+1) \gets {\mathbf{z}^3}(t) + \left({\hat{\mathbf{H}}^3}(t+1)-{\mathbf{H}^3}(t+1)\right),
\end{flalign}
\begin{flalign}\label{equ47}
&{\mathbf{z}^4}(t+1) \gets {\mathbf{z}^4}(t) + \left({\hat{\mathbf{H}}^6}(t+1)-{\mathbf{H}^6}(t+1)\right),
\end{flalign}
\end{subequations}
\end{itemize}

Through a series of iterations, the sequence $(\mathbf{x},\mathbf{y},\Pi,\hat{\Pi})$ converges to a stationary point, and we take the stable point as the solution to the problem based on the below Lemma.
\begin{lemma}\label{lemma33}
For sufficiently large $\rho$, the sequence $(\mathbf{x}^t,\mathbf{y}^t,\Pi^t,\hat{\Pi}^t)$ generated by ADMM algorithm converges to a limit points and all of its limit points are stationary points of the augmented Lagrangian $\mathcal{L}_{\rho}$.
\end{lemma}
\emph{Proof}. Proof can be seen in Appendix B.

Next, we give the detail for the update.

1)~\emph{The solution of the update problem $(\ref{equ40})$}

The problem $(\ref{equ40})$ for updating the introduced variables $\{\mathbf{y}, \hat{\Pi}\}$ can be rewritten into:
\begin{equation*}
\begin{aligned}
\mathcal{P}2.2.1: \mathop{\min}_{\mathbf{y},\hat{\Pi}} \ \ \ &\frac{\rho_0}{2}\sum_{n \in \mathcal{N}}\lVert{\mathbf{x}}(t)-\mathbf{y}^n+\mathbf{\lambda}^{n}(t)\rVert_2^2+\frac{\rho_1}{2}\lVert\hat{\mathbf{R}}^3-{\mathbf{R}^3}(t)+{\mathbf{z}^1}(t)\rVert_2^2+\frac{\rho_2}{2}\lVert\hat{\mathbf{R}}^6-{\mathbf{R}^6}(t)+{\mathbf{z}^2}(t)\rVert_2^2\\
&+\frac{\rho_3}{2}\lVert\hat{\mathbf{H}}^3-{\mathbf{H}^3}(t)+{\mathbf{z}^3}(t)\rVert_2^2+\frac{\rho_4}{2}\lVert\hat{\mathbf{H}}^6-{\mathbf{H}^6}(t)+{\mathbf{z}^4}(t)\rVert_2^2\\
s.t.\ \ \ &\left(\ref{equ49}\right),\left(\ref{equ50}\right),\left(\ref{equ51}\right),\left(\ref{equ52}\right).
\end{aligned}
\end{equation*}

We can see that the above optimization problem can be decomposed into $N_sF$ independent subproblems which are correspond to a request state $n \in \mathcal{N}_{\mathcal{S}}$ for a computation task $f$,
\begin{equation*}
\begin{aligned}
\mathcal{P}2.2.2: &\mathop{\min}_{\{y_{k,f}^n\}_{k=1}^K,\hat{\Pi}_{f,n}}\frac{\rho_0}{2}\sum_{n \in \mathcal{N}_{\mathcal{S}}}\sum_{k \in \mathcal{K}}\left(x_{k,f}(t)-y_{k,f}^n+\lambda_{k,f}^{n}(t)\right)^2+\frac{\rho_1}{2}\left(\hat{R}_{f,n}^3-R_{f,n}^{3}(t)+z_{f,n}^{1}(t)\right)^2\\
&~~~~~~~~~~~~~~+\frac{\rho_2}{2}\left(\hat{R}_{f,n}^6-R_{f,n}^{6}(t)+z_{f,n}^{2}(t)\right)^2+\frac{\rho_3}{2}\left(\hat{H}_{f,n}^3-H_{f,n}^{3}(t)+z_{f,n}^{3}(t)\right)^2\\
&~~~~~~~~~~~~~~+\frac{\rho_4}{2}\left(\hat{H}_f^6-H_{f,n}^{6}(t)+z_{f,n}^{4}(t)\right)^2\\
&~~~~~~s.t. ~~~~\left(\ref{equ49}\right),\left(\ref{equ50}\right),\left(\ref{equ51}\right),\left(\ref{equ52}\right),
\end{aligned}
\end{equation*}
where we use $\hat{\Pi}_{f,n}=\{\hat{R}_{f,n}^3,\hat{R}_{f,n}^6,\hat{H}_{f,n}^3,\hat{H}_{f,n}^6\}$. Obviously, the objective function and the constraints are convex, and the convex programming can be solved efficiently using standard optimization toolbox, e.g., CVX. These  subproblems can be handled in a parallel fashion at different computation units of a centralized controller without effecting the others for saving the computation time.

2)~~\emph{The solution of the update problem $(\ref{equ41})$}

The update problem of the local variables $\{\mathbf{x},\Pi\}$ can be reformulated as
\begin{equation*}
\begin{aligned}
\mathcal{P}2.2.3: \ \ &\mathop{\min}_{\mathbf{x},\Pi} \ \ \ \frac{1}{N_s}\sum_{n \in \mathcal{N}_{\mathcal{S}}}\sum_{f \in \mathcal{F}}\left(R_{f,n}^3\times H_{f,n}^3+R_{f,n}^6\times H_{f,n}^6\right)+\beta\sum_{f \in \mathcal{F}}\sum_{k \in \mathcal{K}}x_{k,f}\left(1-x_{k,f}\right)\\
&~~~~~~~+\frac{\rho_0}{2}\sum_{n \in \mathcal{N}_{\mathcal{S}}}\lVert\mathbf{x}-\mathbf{y}(t+1)+\mathbf{\lambda}(t)\rVert_2^2+\frac{\rho_1}{2}\lVert{\hat{\mathbf{R}}^3}(t+1)-{\mathbf{R}^3}+{\mathbf{z}^1}(t)\rVert_2^2\\
&~~~~~~~+\frac{\rho_2}{2}\lVert{\hat{\mathbf{R}}^6}(t+1)-{\mathbf{R}^6}+{\mathbf{z}^2}(t)\rVert_2^2+\frac{\rho_3}{2}\lVert{\hat{\mathbf{H}}^3}(t+1)-{\mathbf{H}^3}+{\mathbf{z}^3}(t)\rVert_2^2\\
&~~~~~~~ +\frac{\rho_4}{2}\lVert{\hat{\mathbf{H}}^6}(t+1)-{\mathbf{H}^6}+{\mathbf{z}^4}(t)\rVert_2^2\\
& \ \ s.t. ~~\left(\ref{equ5}\right),\\
&\ \ \ \ \ \ \ \mathbf{0}\leqslant\mathbf{x}\leqslant \mathbf{1}.
\end{aligned}
\end{equation*}

For reducing computation complexity, $\mathcal{P}2.2.3$ can also be decomposed into two independent subproblems due to the uncoupled variables $\mathbf{x}$ and $\Pi$. For the computation variable $\mathbf{x}$, the subproblem can be decomposed into $K$ independent subproblems which can be solved parallel. The subproblem for a mobile device $k$ can be written as:
\begin{equation*}
\begin{aligned}
\mathcal{P}2.2.3.1: &\mathop{\min}_{\{x_{k,f}\}_{f=1}^F} \ \beta \sum_{f \in \mathcal{F}}x_{k,f}\left(1-x_{k,f}\right)+\frac{\rho_0}{2}\sum_{n \in \mathcal{N}_{\mathcal{S}}}\sum_{f \in \mathcal{F}}\left(x_{k,f}-y_{k,f}^{n}(t+1)+\lambda_{k,f}^{n}(t)\right)^2\\
& \ \ \ \ s.t. \ \  \left(\ref{equ5}\right),\\
&\ \ \ \ \ \ \ \ \ 0 \leqslant x_{k,f}\leqslant 1,\ \forall f \in \mathcal{F}.
\end{aligned}
\end{equation*}

Similarly, the subproblem for the variables $\Pi$ can be decomposed into $N_sF$ independent problems, and for the sampled request state $n$, the subproblem at the computation task $f$ is
\begin{equation*}
\begin{aligned}
\mathcal{P}2.2.3.2: \mathop{\min}_{\{\Pi_{f,n}\}} \ &\frac{1}{N_s}\left(R_{f,n}^3\times H_{f,n}^3+R_{f,n}^6\times H_{f,n}^6\right)+\frac{\rho_1}{2}\left(\hat{R}_{f,n}^{3}(t+1)-{R_{f,n}^3}+z_{f,n}^{1}(t)\right)^2\\
&+\frac{\rho_2}{2}\left(\hat{R}_{f,n}^{6}(t+1)-{R_{f,n}^6}+z_{f,n}^{2}(t)\right)^2+\frac{\rho_3}{2}\left(\hat{H}_{f,n}^{3}(t+1)-{H_{f,n}^3}+z_{f,n}^{3}(t)\right)^2\\
&+\frac{\rho_4}{2}\left(\hat{H}_{f,n}^{6}(t+1)-{H_{f,n}^6}+z_{f,n}^{4}(t)\right)^2.
\end{aligned}
\end{equation*}

\begin{algorithm}[t]
\renewcommand{\algorithmicrequire}{\textbf{Initialization:}}
\renewcommand{\algorithmicensure}{\textbf{Output:}}
\caption{The corresponding CCCP for $\mathcal{P}2.2.3$}
\label{alg01}
\begin{algorithmic}
	\REQUIRE Decompose problem $\mathcal{P}2.2.3$ into problem $\mathcal{P}2.2.3.1$ and problem $\mathcal{P}2.2.3.2$; Initialize $\{{\mathbf{x}_{k}(1)}, {\Pi_{f,n}(1)}\}=\{{\mathbf{x}_{k}(t)},{\Pi_{f,n}(t)}\} $ and set the number of iteration $i = 1$, the maximum number of iteration $i_{max}$.
	\ENSURE $\{{\mathbf{x}_{k}(t+1)}, {\Pi_{f,n}(t+1)}\}$
\REPEAT
\STATE \emph{1)} Update $\{{\mathbf{x}_{k}(i+1)}, {\Pi_{f,n}(i+1)}\}$ according to $\mathcal{P}2.2.3.3$ and $\mathcal{P}2.2.3.4$;
\STATE \emph{2)} Set $i = i+1$
\UNTIL{convergence of $\{{\mathbf{x}_{k}}, {\Pi_{f,n}}\}$ or $i \geq i_{max}$}.
\end{algorithmic}
\end{algorithm}

It can be found that both $\mathcal{P}2.2.3.1$ and $\mathcal{P}2.2.3.2$ are the DC programming problems with the DC objective function. Therefore, we can adopt successive convex approximation to get a local optimal solution to overcome the difficulty. In the paper, we can solve the problem by using the concave-convex procedure (CCCP) \cite{cccp}. CCCP involves an iterative procedure to solve a sequence of convex subproblems. Specifically, in the $i+1$ iteration, we replace the nonconvex term $R_{f,n}^3\times H_{f,n}^3+R_{f,n}^6\times H_{f,n}^6$ and $x_{k,f}\left(1-x_{k,f}\right)$ by their first-order Taylor expansion:
\begin{equation}\label{equ53}
\begin{aligned}
&R_{f,n}^3\times H_{f,n}^3+R_{f,n}^6\times H_{f,n}^6 = R_{f,n}^{3}(i)\times H_{f,n}^{3}(i)+ R_{f,n}^{6}(i)\times H_{f,n}^{6}(i)+\left[R_{f,n}^{3}(i),H_{f,n}^{3}(i)\right] \cdot \\
&\left[H_{f,n}^3- H_{f,n}^{3}(i),R_{f,n}^3-R_{f,n}^{3}(i)\right]^{T}+\left[R_{f,n}^{6}(i),H_{f,n}^{6}(i)\right] \cdot \left[H_{f,n}^6 - H_{f,n}^{6}(i),R_{f,n}^6-R_{f,n}^{6}(i)\right]^{T}\\
&~~~~~~~~~~~~~~~~~~~~~~~~~~~~~~~= R_{f,n}^{3}(i)\times H_{f,n}^3 + R_{f,n}^3\times H_{f,n}^{3}(i)-R_{f,n}^{3}(i)\times H_{f,n}^{3}(i)+\\
&R_{f,n}^{6}(i)\times H_{f,n}^6 + R_{f,n}^6\times H_{f,n}^{6}(i)-R_{f,n}^{6}(i)\times H_{f,n}^{6}(i).
\end{aligned}
\end{equation}
\begin{equation}\label{equ54}
x_{k,f}\left(1-x_{k,f}\right) = x_{k,f}-2{x_{k,f}}(i)x_{k,f}+{x_{k,f}}(i){x_{k,f}}(i).
\end{equation}

\begin{algorithm}[t]
\renewcommand{\algorithmicrequire}{\textbf{Initialization:}}
\renewcommand{\algorithmicensure}{\textbf{Iteration:}}
\caption{The ADMM-Based Method for $\mathcal{P}2$}
\label{alg02}
\begin{algorithmic}
	\REQUIRE Initialize the number of iteration $t = 1$, the maximum number of iteration $t_{max}$, the error limit $\epsilon = 1$, the penalty parameter $\beta=100$, the dual variables $\mathbf{\lambda}$ and $\mathbf{z}$ are initialized as $\mathbf{0}$.
	\ENSURE
\REPEAT
\STATE \emph{1)} Update $\{\mathbf{y}, \hat{\Pi}\}$ based on the subproblems $\mathcal{P}2.2.2$ for each computation task $f$ at a sampled request state $n$;
\STATE \emph{2)} Update $\{\mathbf{x},\Pi\}$ rely on the decomposed subproblem $\mathcal{P}2.2.3$ solved by Algorithm $\ref{alg01}$;
\STATE \emph{3)} Update $\{\mathbf{\lambda}, \mathbf{z}\}$ according to $(\ref{equ43})$, $(\ref{equ44})$, $(\ref{equ45})$, $(\ref{equ46})$, $(\ref{equ47})$;
\STATE \emph{4)} $t = t+1$
\UNTIL{$\lVert\mathbf{x}(t+1)-\mathbf{x}(t)\rVert \leqslant \epsilon$ or $t \geqslant t_{max}$}.
\end{algorithmic}
\end{algorithm}

For the $i+1$ iteration in CCCP, $\mathcal{P}2.2.3.1$ and $\mathcal{P}2.2.3.2$ are transformed into the following programming
\begin{align*}
\mathcal{P}2.2.3.3:  \mathop{\min}_{\{x_{k,f}\}_{f=1}^F} &\beta\sum_{f \in \mathcal{F}}\left(x_{k,f}-2{x_{k,f}(i)}x_{k,f}\right)+\frac{\rho_0}{2}\sum_{n \in \mathcal{N}_{\mathcal{S}}}\sum_{f \in \mathcal{F}}\left(x_{k,f}-y_{k,f}^{n}(t+1)+{\lambda_{k,f}^n}(t)\right)^2\\
\ \ \   s.t.~~~&\left(\ref{equ5}\right),\\
&0 \leqslant x_{k,f}\leqslant 1,\forall f \in \mathcal{F}.
\end{align*}
\begin{align*}
\mathcal{P}2.2.3.4:  &\mathop{\min}_{\{\Pi_{f,n}\}} \frac{1}{N_s}\sum_{f \in \mathcal{F}}\left(R_{f,n}^{3}(i) \times H_{f,n}^3 + R_{f,n}^3 \times H_{f,n}^{3}(i)+R_{f,n}^{6}(i)\times H_{f,n}^6 + R_{f,n}^6\times H_{f,n}^{6}(i)\right)\\
&+\frac{\rho_1}{2}\sum_{f \in \mathcal{F}}\left(\hat{R}_{f,n}^{3}(t+1)-{R_{f,n}^3}+z_{f,n}^{1}(t)\right)^2+\frac{\rho_2}{2}\sum_{f \in \mathcal{F}}\left(\hat{R}_{f,n}^{6}(t+1)-R_{f,n}^6+z_{f,n}^{2}(t)\right)^2\\
&+\frac{\rho_3}{2}\sum_{f \in \mathcal{F}}\left(\hat{H}^{3}(t+1)-{H_{f,n}^3}+z_{f,n}^{3}(t)\right)^2+\frac{\rho_4}{2}\sum_{f \in \mathcal{F}}\left(\hat{H}_{f,n}^{6}(t+1)-{H_{f,n}^6}+z_{f,n}^{4}(t)\right)^2.
\end{align*}

$\mathcal{P}2.2.3.3$ and $\mathcal{P}2.2.3.4$ are convex problems and thus both can be solved efficiently by the standard convex optimization toolbox. Then the near-optimal solution $\{{\mathbf{x}_{k}(t+1)}, {\Pi_{f,n}(t+1)}\}$ is obtained by iteratively solving $\mathcal{P}2.2.3.1$ and $\mathcal{P}2.2.3.2$ until it converges as shown in the above Algorithm $\ref{alg01}$.

In summary, the effective computation scheme $\mathbf{x}^{\ast}$ is obtained by leveraging the ADMM algorithm to $\mathcal{P}2$. Through successive iterations, we firstly minimize the augmented Lagrangian function $(\ref{equl})$ over the introduced variables and decompose the optimization problem into smaller subproblems, which is executed in parallel to improve the computation speed. Next, we update the global variables based on the problem $(\ref{equ40})$. Similarly, the problem is decomposed into several subproblems for reducing computation time, and CCCP is utilized to obtain the  solution for each subproblem. Finally, the dual variables are updated based on $(\ref{equ43})$, $(\ref{equ44})$, $(\ref{equ45})$, $(\ref{equ46})$, $(\ref{equ47})$. The method is summarized in Algorithm $\ref{alg02}$

\subsection{Algorithm to Solve Problem $\mathcal{P}3$}

$\mathcal{P}3.1$ is formulated based on $d=1$, while $\mathcal{P}3.2$ is expressed for $d=0$. Both of the two subproblems are nonsmooth and nonconvex 0-1 programming because of $(\ref{equ8})$, $(\ref{equ14})$. It is unreasonable to relaxing the binary constraints to continuous constraints which approximates the nonsmooth 0-1 programming with a smooth one as Lemma $\ref{lem02}$ due to the special structure of $b^q\left(\mathbf{c},\mathbf{x}\right)$ in Proposition $\ref{lem01}$. Obviously, the computational complexity produced by traversing each feasible solution can reach $\mathcal{O}(2^F)$ which is pretty high especially in the large scale programming. Therefore, we propose an algorithm to obtain the acceptable solution of problem $\mathcal{P}3.1$ and $\mathcal{P}3.2$ as shown in the following Algorithm $\ref{alg03}$.

\begin{algorithm}
\renewcommand{\algorithmicrequire}{\textbf{Input:}}
\renewcommand{\algorithmicensure}{\textbf{Initialization:}}
\caption{The Proposed Algorithm for $\mathcal{P}3.1$, $\mathcal{P}3.2$}
\label{alg03}
\begin{algorithmic}
\REQUIRE
The sampling set of request state $\mathcal{N}_{\mathcal{S}}$;
The computation scheme $\mathbf{x}^{\ast}$;
The set of positive computational task $\hat{\mathcal{F}}$ ($\forall f \in \mathcal{F}$, for $\mathcal{P}3.1$, $\exists k \in \mathcal{K}$, $x_{k,f}^{\ast}=1$; for $\mathcal{P}3.2$, $\exists k \in \mathcal{K}$, $x_{k,f}^{\ast}=0$);
\ENSURE
The initial number of cached computation task $\mathnormal{num}_{max} = |\hat{\mathcal{F}}|$;
The initial parameter $t_{0}$(for $\mathcal{P}3.1$,$t_{0} =  \lfloor \frac{CK}{\mathnormal{num}_{max}I} \rfloor$ and $t_{0} =  \lfloor \frac{CK}{\mathnormal{num}_{max}O} \rfloor$otherwise);
The set $\hat{\mathcal{F}}$ is sort by $\sum_{n \in \mathcal{N}_{\mathcal{S}}}\sum_{k \in \mathcal{K}}x_{k,f}^{\ast}s_{k,f}^n$ in descending order based on $\mathcal{P}3.1$, and $\sum_{q \in \mathcal{N}_{\mathcal{S}}}\sum_{k \in \mathcal{K}}(1-x_{k,f}^{\ast})s_{k,f}^n$ is the standard of the descending order according to $\mathcal{P}3.2$;
The initial cache scheme ${\textbf{c}}^{\ast} \triangleq ({c}^{\ast}_{f})_{f\in \mathcal{F}}$, ${c}^{\ast}_{f} = 1$, for $f \in \hat{\mathcal{F}}$, and ${c}^{\ast}_{f} = 0$ otherwise;
The corresponding average bandwidth ${\mathbf{B}}^{\ast}$ based on $\mathbf{x}^{\ast}$ and ${\mathbf{c}}^{\ast}$;
The initial variable $\mathnormal{num} = \mathnormal{num}_{max} - 1$;

\WHILE {$\mathnormal{num} \geqslant 0$}
\STATE $t = \lfloor \frac{CK}{\mathnormal{num}I} \rfloor$ for $\mathcal{P}3.1$,or $t = \lfloor \frac{CK}{\mathnormal{num}O} \rfloor$ for $\mathcal{P}3.2$;

\IF {$t_{0} \ne t$}
\STATE \emph{1)} The first $\mathnormal{num}$ task in $\hat{\mathcal{F}}$ are decided to be cached, and the cache scheme is defined as $\mathbf{c}$;
\STATE \emph{2)} The corresponding average bandwidth $\mathbf{B}$ based on $\mathbf{x}^{\ast}$ and $\mathbf{c}$;
\STATE \emph{3)} Compare the average bandwidth $\mathbf{B}$ and ${\mathbf{B}}^{\ast}$;

\IF {$\mathbf{B} \leqslant {\mathbf{B}}^{\ast}$}
       \STATE ${\mathbf{B}}^{\ast}=\mathbf{B}$ and $\mathbf{c}^{\ast}=\mathbf{c}$;
\ENDIF
\STATE $t_{0} = t$;
\ENDIF
\STATE $\mathnormal{num}=\mathnormal{num}-1$;
\ENDWHILE
\RETURN ${\mathbf{B}}^{\ast}$, $\mathbf{c}^{\ast}$ of $\mathcal{P}3.1$, $\mathcal{P}3.2$.
\end{algorithmic}
\end{algorithm}

The algorithm provides a new search method for problem $\mathcal{P}3.1$, $\mathcal{P}3.2$. The storage of each MD is full exploited, that is the number of tasks coded cached is the maximum number of tasks meet a certain $t$. What's more, we choose the coded caching scheme according to the number of tasks requested for each comparison in a iteration which maintains the global gain of the coded caching scheme as well as reduces the computation complexity. After applying the algorithm for solving problem $\mathcal{P}3.1$ and $\mathcal{P}3.2$ respectively, the cache scheme $d^{\ast}$ and $\mathbf{c}^{\ast}$ is obtained by comparing ${\mathbf{B}^{\ast}}$.

\subsection{Analysis of Algorithms Proposed}
For obtaining the computation scheme $\mathbf{x}^{\ast}$, the computation complexity of solving $(\mathcal{P}2.2.2)$ is $\mathcal{O}(K)$ during each iteration in ADMM algorithm. The update problem of the variables $\{\mathbf{y}, \hat{\Pi}\}$ need to solve $N_sF$ subproblems, and the total computational complexity is $\mathcal{O}(N_sKF)$. Similarly, the computational complexity of updating the variables $\{\mathbf{x}, \Pi\}$ can be expressed as $\mathcal{O}(K+N_sF)$ in a CCCP iteration. Supposed that the numbers of iterations required by the ADMM algorithm and the CCCP algorithm are $N_{ADMM}$ and $N_{CCCP}$, respectively. Therefore, the computational complexity of the proposed algorithm in the subsection is formulated as $\mathcal{O}(N_{ADMM}(N_sKF+N_{CCCP}(K+N_sF)))$. On the other hand, the maximum computational complexity produced by the proposed algorithm in the paper can be achieved $\mathcal{O}(2K)$, which is far less than traversing each feasible solution for the coded caching decision $\mathbf{c}$, $d$.

In a word, the computation scheme $\mathbf{x}$ and the coded cache scheme $\mathbf{c}$, $d$ are decoupled in the paper, then the original problem $\mathcal{P}1$ is decomposed into three subproblems. The suboptimal solution $\mathbf{x}^{\ast}$ is obtained through Algorithm $\ref{alg02}$ where we utility ADMM algorithm and the update problem is decomposed into several subproblems for parallel computation in each iteration. The computation complexity is reduced and the computation speed is increased. Moreover, we propose Algorithm $\ref{alg03}$ to get the acceptable solution $\mathbf{c}^{\ast}$ and $d^{\ast}$ for reducing the computation complexity compared with the traditional algorithm.
\section{Simulation}
In this section, numerical results are provided to validate the effectiveness of the proposed scheme. Without loss of generality, the input data size and the output data size of each task are $3$ Mbits and $6$ Mbits in the simulation \cite{TCOM_sun}. The probability of each MD requests the task is identical and independent, as well as the samples $N_{s}$ is set to be 1000. The average energy $E_k$ for each MD is uniformly assigned from the set $[0,150]$ $J$. The computation load $w_f$ of the task $f$ follows a uniform distribution in the range $[5,10]$ cycles per bit, and the latency is set to be $20$ ms. The channel $h_k$ is modelled as Rayleigh fading, i.e., $h_k\sim CN(0,1)$, and the average signal-to-noise (SNR) is uniformly selected from the set $[10, 20]$ dB for different mobile devices. The constant $\alpha$ for computing the average energy is $10^{-24}$.

We compare the proposed scheme with the following three benchmarks:
\begin{itemize}
\item \textbf{Local Coded Cache}: The policy takes only the cache capability into consideration, that is, all computation tasks are processed by the MEC server, and the entire output data of a task or the coded output data is transmitted to MDs which only stores the output data of the tasks.
  \item \textbf{Local Computing}: The computation capability of MDs is considered only in the case. The tasks can be computed locally, or by the MEC server. Thus, the entire input data or the entire output data of a task is delivered to MDs.
  \item \textbf{Traditional Transmission}: In the case, BS only multicasts the entire output data of a task to MDs, without using the caching and computing capability of MDs.
\end{itemize}

%
%
%

\begin{figure}[t]
\centering
\begin{minipage}[c]{0.45\textwidth}
\centering
\includegraphics[width=7.5cm]{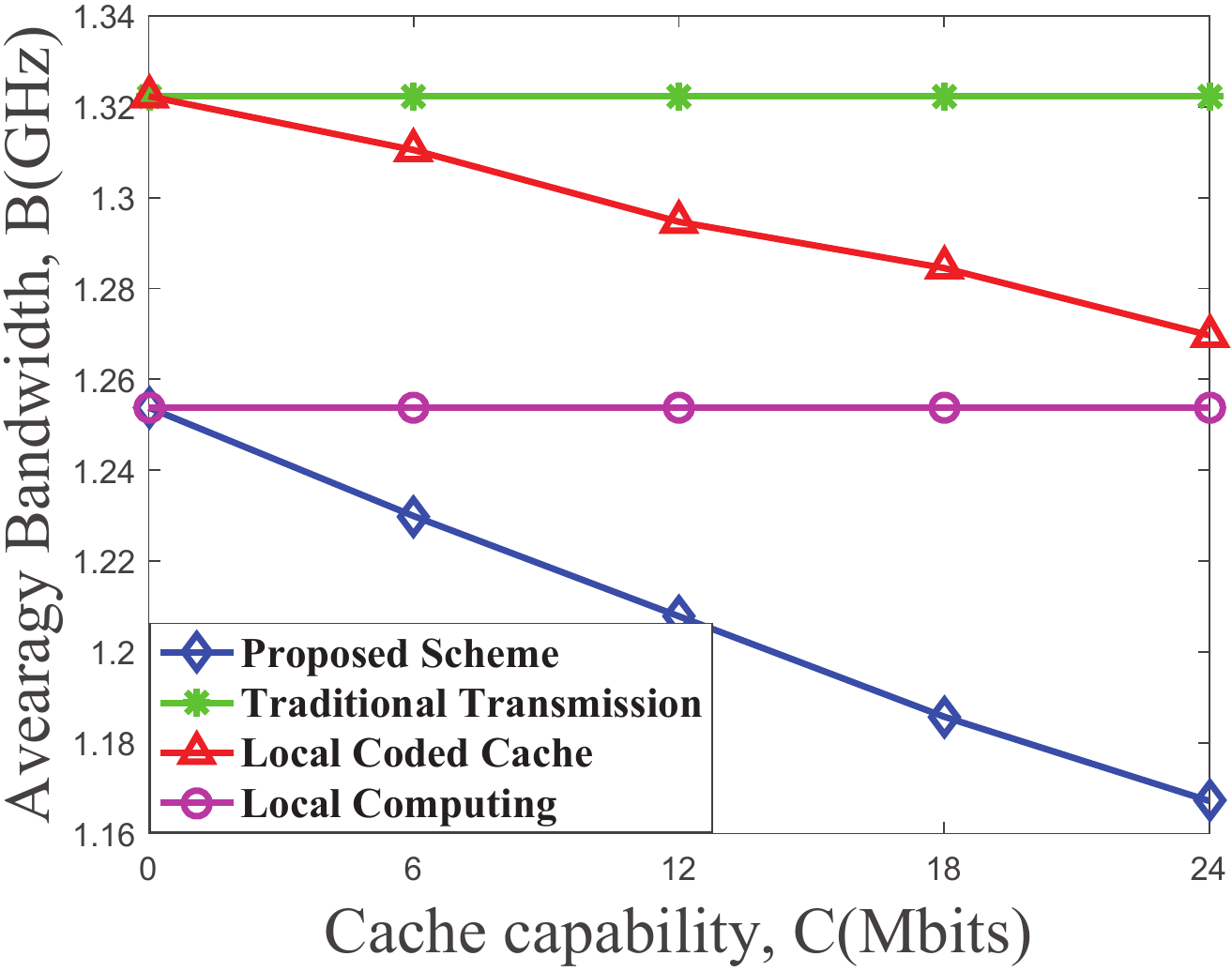}
\caption{Impact of cache capability capability.}
\label{simulation}
\end{minipage}%
\begin{minipage}[c]{0.45\textwidth}
\centering
\includegraphics[width=7.5cm]{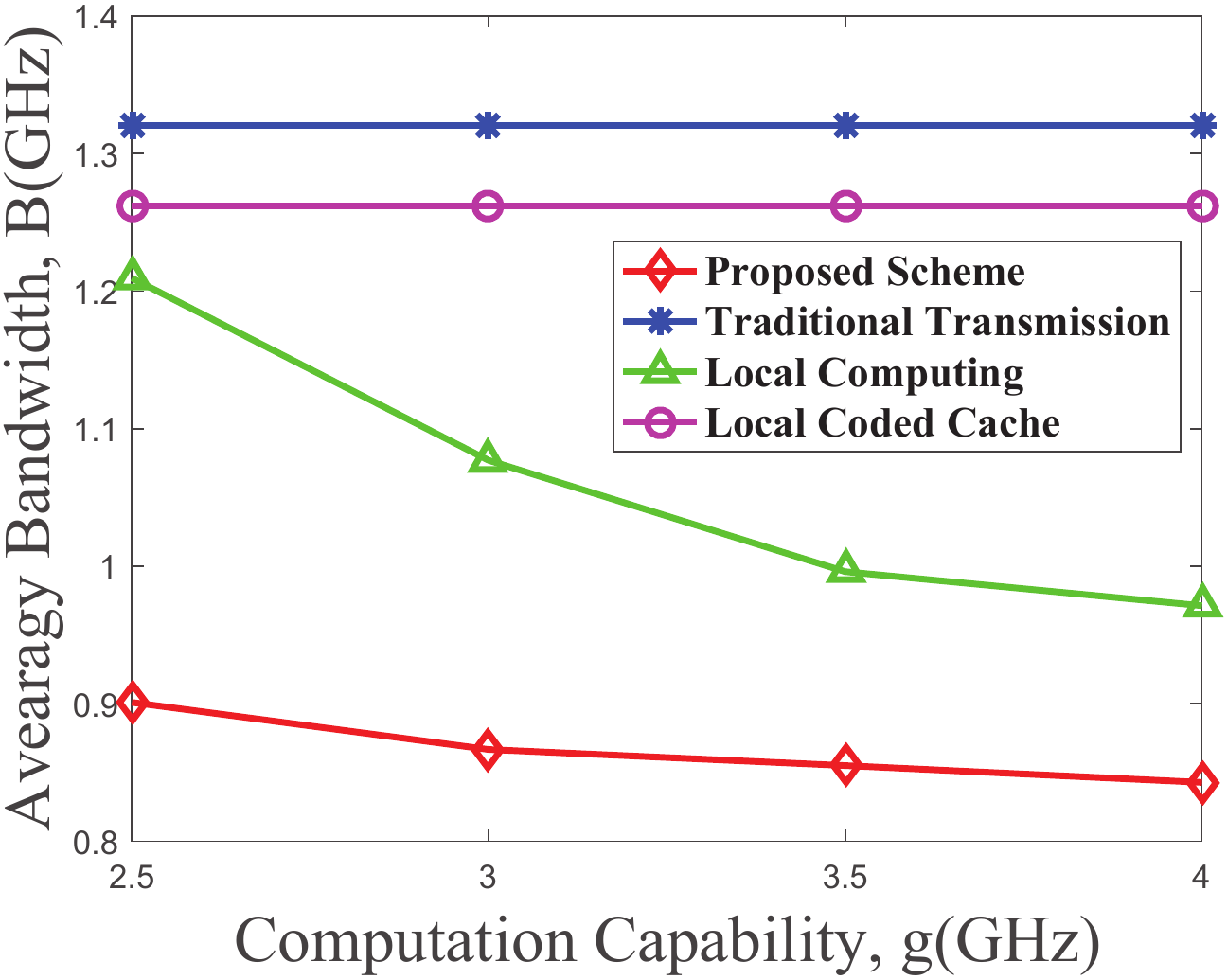}
\caption{Impact of computation capability.}
\label{simulation00}
\end{minipage}
\end{figure}

\begin{figure}[t]
\centering
\begin{minipage}[c]{0.45\textwidth}
\centering
\includegraphics[width=7.5cm]{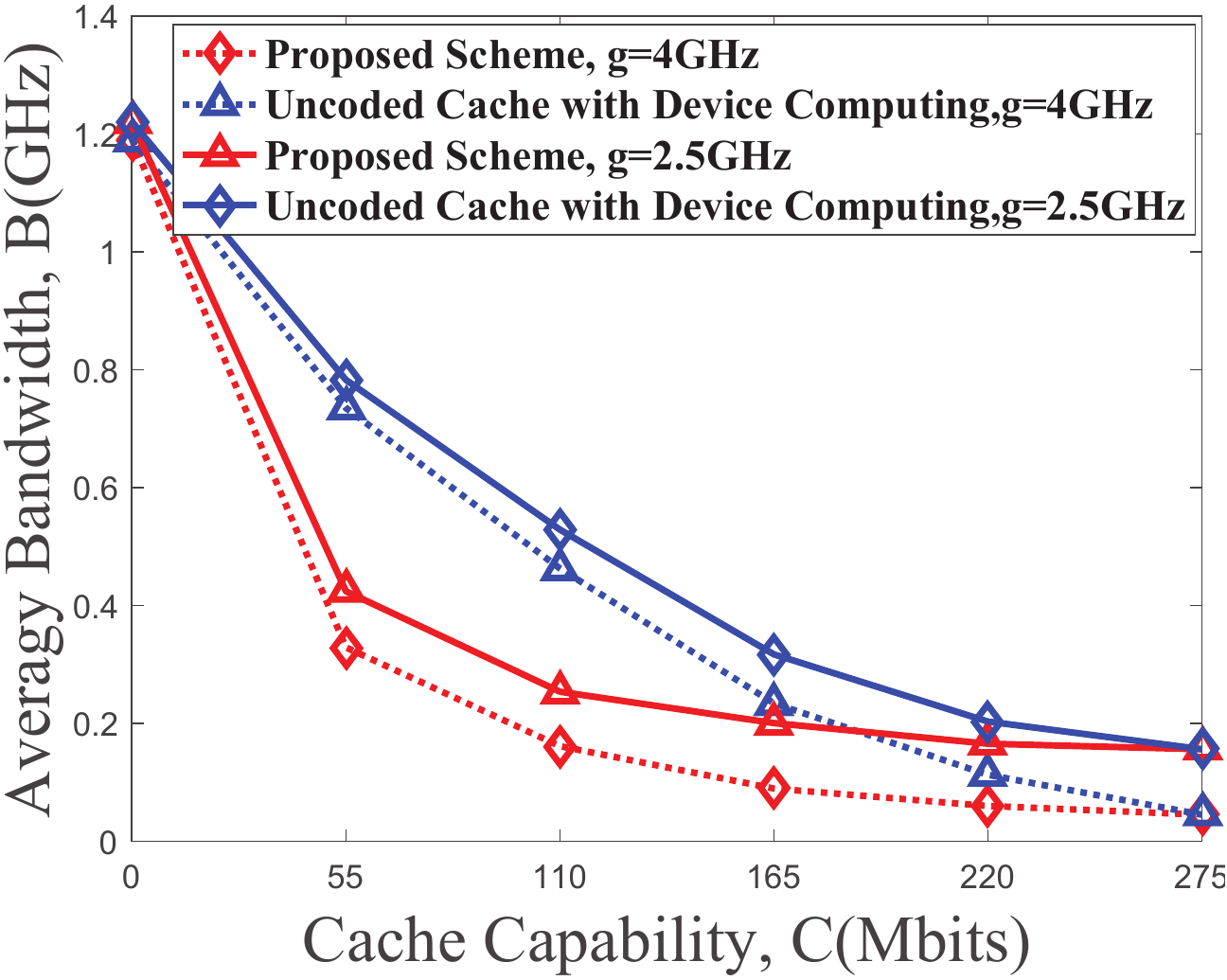}
\caption{Impact of the coded cache strategy}
\label{simulation01}
\end{minipage}
\begin{minipage}[c]{0.45\textwidth}
\centering
\includegraphics[width=7.5cm]{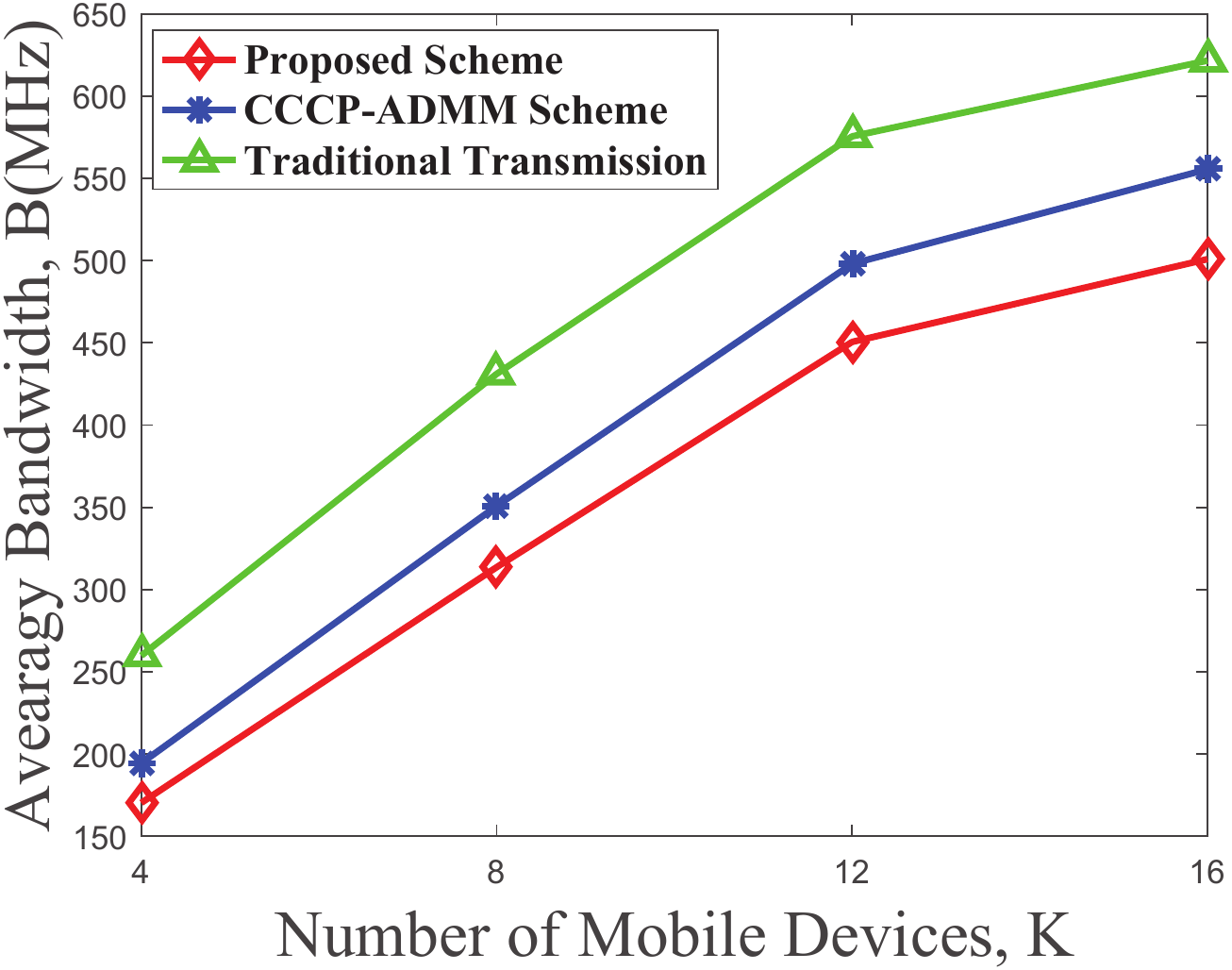}
\caption{Impact of the proposed algorithm}
\label{simulation02}
\end{minipage}
\end{figure}

Fig.~\ref{simulation} and Fig.~\ref{simulation00} illustrate the impact of cache size and computing capability on the average bandwidth cost where there are $K=20$ MDs and $F=100$ computation tasks. Intuitively, observe that the average bandwidth monotonously decreases with $C$ and $g$. It shows the proposed scheme achieves minimum bandwidth consumption over the baselines by making full use of the caching and computing resources of the mobile devices. We also can see from Fig.~\ref{simulation} that even if the mobile device does not have the caching ability, i.e., $C=0$, the proposed scheme also can save the bandwidth compared with the traditional transmission scheme by using the local computing. This suggestion is also depicted in Fig.~\ref{simulation00}, e.g., the proposed scheme has a significant performance gain compared with the local coded cache scheme. For example, the required bandwidth can reduce from $1.31$ GHz (the local coded cache scheme) to $0.87$ GHz (the proposed scheme) at $g=3.5$ GHz.

What's more, we observe from Fig.~\ref{simulation} that the performance gap between the proposed scheme and the local computing scheme is larger than that between the tradition transmission scheme and the local coded cache scheme, which means the computing resource in MDs can help the system to achieve more caching gain than that of without exploiting the computing resource. Similarly, the cache resource of MDs can bring more gain by comparing the gap between the proposed scheme and the local coded cache scheme with the gap between the tradition transmission scheme and the local computing scheme Fig.~\ref{simulation00}.

Next, similar to \cite{coded}, Fig.~\ref{simulation01} evaluates the coded gain of the proposed scheme, compared with the uncoded caching with device computing scheme which the entire input data or the entire output data of a task is cached in a MD. The coded caching with device computing, i.e., the proposed scheme of this paper, can bring significant coded gain over the uncoded caching with device computing scheme, e.g., reducing the bandwidth from $450$ MHz to $190$ MHz with $C=110$ Mbits and $g=4$ GHz.

Finally, Fig.~\ref{simulation02} verifies the effectiveness of the proposed algorithm by comparing with the CCCP-ADMM algorithm used in \cite{TWC_sun}. We can observe that the proposed algorithm still achieves good performance gains over the CCCP-ADMM algorithm, especially when the number of mobile devices is large (e.g.,$K \geqslant 12$). This is because the sufficient conditions for ADMM to converge on monotonic programs hold in our optimization problem, we can directly obtain the stationary point from the ADMM algorithm, as shown in Lamma \ref{lemma33}, while CCCP converges to a local minimum in the CCCP-ADMM algorithm.
\section{Conclution}
In the paper, we have studied the problem of how to save the average transmission bandwidth by exploiting the caching and computing resources of MDs in the MEC system. A coded caching with device computing strategy is proposed to minimize the average bandwidth under the delay of the computation tasks, the cache size and the average energy consumption of MDs. The formulated problem is a large-scale mix integer nonconvex and nonsmooth programming when the numbers of MDs and computation tasks get larger. Obviously, the programming in the paper is difficult to be solved and thus we have decoupled it into several subproblems which can be solved separately in an efficient way. The numerical results show that the coded cache with device computing scheme significantly outperforms the three state-of-the-art benchmarks.

\begin{appendices}
\section{Proof of Proposition 1}
Based on the computation scheme $\mathbf{x}$ and the cached scheme $\mathbf{c}$, $d$, the integer $t$ can be obtained. Obviously, the computation task $f$ is cached on all of the mobile devices when $c_{f}=1$, and the task is not split if $t = K$ where the integral tasks are cached and there is no data need to be transmitted. Therefore we draw a conclusion that $b^q\left(\mathbf{c},\mathbf{x}\right)=0$ when $t=K$.

If $t<K$, then all cached tasks are split into $\binom{K}{t}$ nonoverlapping subfiles of equal size. The total rate of coded multicast transmission $b^q\left(\mathbf{c},\mathbf{x}\right)$ depends on the set of mobile devices request coded multicast transmission $\mathcal{M}^q$ with $\lvert \mathcal{M}^q \rvert = M^q$. There is no mobile device that requests the coded cached task when $M^q=0$, and the data size of coded multicast transmission is zero, i.e., $b^q\left(\mathbf{c},\mathbf{x}\right)=0$. For each subset $\mathcal{V}^q \subset \mathcal{K}$ of cardinality $\lvert \mathcal{V}^q \rvert = t+1$, the coded multicast transmission for a subset $\mathcal{V}^q$ is
\begin{equation}
\oplus_{m^q \in \mathcal{V}}W_{d_{m^q}, \mathcal{V}\backslash \{m^q\}}.
\end{equation}
The data size rate of the transmission is $1/\binom{K}{t}$. The number of subsets $\mathcal{V}^q$ which contains one of the mobile device element $m_{1}^q$ is $\binom{K-1}{t}$, $m_{1}^q \in \mathcal{M}^q$. Then, there are $\binom{K-2}{t}$ subsets $\mathcal{V}^q$ that includes the element $m_{2}^q$ without $m_{1}^q$ suppose that $m_{2}^q \in \mathcal{M}^q$ and $m_{1}^q \ne m_{2}^q$. We can reduce the rest from the above analogy that the total number of the subsets where each subset contents at least an element $m^q$, $m^q \in \mathcal{M}^q$ is $\sum_{i=1}^{M^q} \binom{K-i}{t}$. Thus the rate of coded multicast transmission is
\begin{equation}
b^q\left(\mathbf{c},\mathbf{x}\right) = \sum_{i=1}^{M^q} \binom{K-i}{t}/\binom{K}{t}.
\end{equation}

Obviously, each subset that satisfies $\mathcal{V}^q \subset \mathcal{K}:\lvert \mathcal{V}^q \rvert = t+1$ includes at least an element $m^q \in \mathcal{M}^q$ when $M^q > K-t$. Therefore, the number of the subsets is $\binom{K}{t+1}$. Similarly, the rate can be expressed as
\begin{equation}
b^q\left(\mathbf{c},\mathbf{x}\right) = \binom{K}{t+1}/{\binom{K}{t}}=\frac{K-t}{1+t}.
\end{equation}

\section{proof of lemma 3}
By introducing the variables $\{\mathbf{y},\hat{\Pi}\}$, we know that $\{\mathbf{x},\Pi\}$ and $\{\mathbf{y},\hat{\Pi}\}$ are decoupled in the objective function. The objective function $\Psi$ in problem $\mathcal{P}2.2$ is coercive over the set, that is, $\Psi \to \infty$ if the variables $\{\mathbf{x},\mathbf{y},\Pi,\hat{\Pi}\} \in \mathcal{C}$ and $\Arrowvert\{\mathbf{x},\mathbf{y},\Pi,\hat{\Pi}\} \Arrowvert \to \infty$ where $\mathcal{C}$ is the feasible set for those variables.

We can obtain that $\mathbf{x}-\mathbf{y}^n=\mathbf{0}$, for all $n \in \mathcal{N}_{\mathcal{S}}$. Based on \cite{admm01}, $\mathrm{A}\mathbf{x}+\mathrm{B}_n\mathbf{y}^n=\mathbf{0}$. Therefore, the matrix $\mathrm{A}$ is $\textbf{I}_{K \times K}$ and $\mathrm{B}_n=-\textbf{I}_{K \times K}$. $\mathrm{Im}(\mathrm{A})$ is the set of all vectors in $\mathbb{R}^{K}$ due to the ranks of the matrix are $K$ unit vectors $\mathbf{v}_{i}$ ($i=1,2,\dots,K$), and $\mathrm{Im}(\mathrm{B}_n)$ equals the set of all vectors in $\mathbb{R}^{K}$. Finally, $\mathrm{Im}(\mathrm{A}) = \mathrm{Im}(\mathrm{B}^n)$. Similarly, $\Pi-\hat{\Pi}=\mathbf{0}$, and the matrices $\mathrm{C}$ is $\textbf{I}_{N_{s} \times N_{s}}$ and $\mathrm{D} =-\textbf{I}_{N_{s} \times N_{s}}$. What's more, $\mathrm{Im}(\mathrm{C}) = \mathrm{Im}(\mathrm{D})$, and those are both the set of all vectors in $\mathbb{R}^{N_s}$. The matrices $\mathrm{A}$, $\mathrm{B}_n$, $\mathrm{C}$ and $\mathrm{D}$ are full column rank, their null spaces are trivial and, thus, the unique minimizer $\mathrm{H}$ for any fixed $\{\mathbf{x},\Pi\}$ and $\mathrm{F}$ that is the minimizer for those fixed $\{\mathbf{y},\hat{\Pi}\}$ reduce to linear operators are Lipschitz continuous map.
The constraints are convex and continuous in addition to the objective function. The objective function and the derivative of the objective function are wrote as below,
\begin{align*}
h(\mathbf{x}, \Pi)&=h_1(\Pi)+h_2(\mathbf{x})\\
&=\sum_{n \in \mathcal{N}}\sum_{f \in \mathcal{F}}\left(R_{f,n}^I\times H_{f,n}^I+R_{f,n}^O\times H_{f,n}^O\right)+\beta\sum_{f \in \mathcal{F}}\sum_{k \in \mathcal{K}}x_{k,f}\left(1-x_{k,f}\right),
\end{align*}
where the variable $\Pi$ has been given, $\Pi_0=\{\mathbf{R}_0^I, \mathbf{R}_0^O,\mathbf{H}_0^I, \mathbf{H}_0^O\}$.
\begin{subequations}
\begin{flalign}
&\frac{\partial h_1(\Pi)}{\partial(\mathbf{R}^I)}=\mathbf{H}_0^I, \ \  \frac{\partial h_1(\Pi)}{\partial(\mathbf{H}^I)}=\mathbf{R}_0^I, \ \  \frac{\partial h_1(\Pi)}{\partial(\mathbf{R}^O)}=\mathbf{H}_0^O, \ \ \frac{\partial h_1(\Pi)}{\partial(\mathbf{H}^O)}=\mathbf{R}_0^O,
\end{flalign}
\begin{flalign}
&\frac{\partial h_2(\mathbf{x})}{\partial(\mathbf{x})}=1-2\mathbf{x}.
\end{flalign}
\end{subequations}

Note that the objective function is differentiable and the derivative of  are globally Lipschitz continuous to each variables with constant $\Pi_0$, and -2. Therefore, the function $h(\mathbf{x}, \Pi)$ is Lipschitz differentiable.

The indicator functions based on the constraints \eqref{equ5}, \eqref{equ49}, \eqref{equ50}, \eqref{equ51},\eqref{equ52} are lower semi-continuous since the feasible sets are convex and closed. Therefore, we can get the lemma 3 based on Theorem 1 in \cite{admm01}, that is, the sequence $(\mathbf{x}^t,\mathbf{y}^t,\Pi^t,\hat{\Pi}^t,\lambda^t,\mathbf{z}^t)$ obtained by ADMM algorithm has limit points and all of its limit points are stationary are stationary points of the augmented Lagrangian $\mathcal{L}_{\rho}$ for any sufficiently large $\rho$.

\end{appendices}
\bibliographystyle{IEEEtran}
\bibliography{VOD_paper_v1}

\end{document}